\documentclass[useAMS,usenatbib]{mn2e}

\include{epsf}
\usepackage{multirow}
\usepackage{graphicx}
\usepackage{amssymb}

\title[QSO-LRG 2-Point Cross-Correlation Function and Redshift-Space
 Distortions]{QSO-LRG 2-Point Cross-Correlation Function and Redshift-Space
 Distortions}

\author[Mountrichas et al.]{G. Mountrichas,$^1$\thanks{E-mail:
 georgios.mountrichas@durham.ac.uk}, T. Shanks$^1$, S. M. Croom$^2$, U. Sawangwit$^1$, D. P. Schneider$^3$, \newauthor
    A. D. Myers$^4$, K. Pimbblet$^5$\\ 
$^1$Department of Physics, University of Durham, South Road, Durham DH1 3LE, UK\\
$^2$School of Physics, University of Sydney, NSW 2006 Australia\\
$^3$Department of Astronomy and Astrophyscics, Penn State, 504 Davey Laboratory, University Park, Pennsylvania 16802\\
$^4$Department of Astronomy, MC-221, University of Illinois, 1002 W. Green Street, Urbana, IL 61801 USA\\
$^5$Department of Physics, University of Queensland, Brisbane, Qld 4072, Australia}

\begin{document}

\date{27 March 2007}

\pagerange{\pageref{firstpage}--\pageref{lastpage}} \pubyear{2007}

\maketitle

\label{firstpage}

\begin{abstract}

We have measured the bias of QSOs as a function of QSO luminosity at
fixed redshift ($z<1$) by cross-correlating them with Luminous Red
Galaxies (LRGs) in the same spatial volume, hence breaking the
degeneracy between QSO luminosity and redshift. We use three QSO samples
from 2SLAQ, 2QZ and SDSS covering a QSO absolute magnitude range,
$-24.5<M_{b_J}<-21.5$, and cross-correlate them with 2SLAQ
($z\approx0.5$) and AAOmega ($z\approx0.7$) photometric and
spectroscopic LRGs in the same redshift ranges. The spectroscopic QSO
samples, in the spectroscopic LRG areas, contain 300-700 QSOs and in
photometric LRG areas up to 7000 QSOs. The 2-D and 3-D cross-clustering
measurements are generally in good agreement.  Our (2SLAQ) QSO-LRG
clustering amplitude ($r_0=6.8_{-0.3}^{+0.1}$h$^{-1}$Mpc) as measured
from the semi-projected cross-correlation function appears similar to
the (2SLAQ) LRG-LRG auto-correlation amplitude
($r_0=7.45\pm0.35$h$^{-1}$Mpc) and both are higher than the (2QZ+2SLAQ)
QSO-QSO amplitude ($r_0\simeq5.0$h$^{-1}$Mpc). Our measurements show
remarkably little QSO-LRG cross-clustering dependence on QSO
luminosity. If anything, the results imply that brighter QSOs may be
less highly biased than faint QSOs, the opposite direction expected from
simple high peaks biasing models, where  more luminous QSOs are assumed
to occupy rarer high mass peaks. Assuming a standard $\Lambda$CDM model
and values for $b_{LRG}$ measured from LRG autocorrelation analyses, we
find $b_Q=1.45\pm0.11$ at $M_{b_J}\approx-24$ and $b_Q=1.90\pm0.16$ at $M_{b_J}\approx-22$.

We also find consistent results for the QSO bias from a $z-$space
distortion analysis of the QSO-LRG cross-clustering at $z\approx0.55$.
The velocity dispersions fitted to QSO-LRG cross-correlation, $\xi(\sigma,\pi)$, at $\pm680$ kms$^{-1}$ are intermediate between those for
QSO-QSO and LRG-LRG clustering, as expected given the larger QSO redshift
errors. The dynamical infall results give $\beta _Q=0.55\pm0.10$, implying
$b_Q=1.4\pm0.2$. Thus both the $z-$space distortion and the amplitude
analyses yield $b_Q\approx1.5$ at $M_{b_J}\approx-23$. The implied dark matter
halo mass inhabited by QSOs at $z\approx0.55$ is
$\sim10^{13}h^{-1}M_{\sun}$, again approximately independent of QSO luminosity.

\end{abstract}

\begin{keywords}
galaxies:clusters:general-quasars:general-cosmology:observations-large-scale structure of Universe
\end{keywords}

\section{Introduction}

Useful constraints on the nature of QSOs can be drawn from even the
simplest measure of QSO clustering, the amplitude of the real-space
2-point correlation function. For example, measuring QSO clustering on
small scales can constrain models of galaxy mergers and quasar formation
(Myers et al. 2007). Furthermore, the redshift evolution and luminosity
dependence of the QSO bias can be studied. Recent work on the clustering
dependence of the QSO bias on redshift suggested evolution of the 2QZ
QSO bias (Croom et al. 2005). However, the fact that the most luminous
QSOs lie at high redshifts, while the faintest are at low redshifts,
made it difficult to study how QSO properties depend on luminosity. Now,
surveys such as 2SLAQ (Cannon et al. 2006) that span a wide range of QSO luminosities, have
broken that redshift-luminosity degeneracy and revealed little QSO
clustering dependence on luminosity, at fixed redshift (da
$\hat{A}$ngela et al. 2006). Moreover, the amplitude of the QSO
clustering is correlated with the average mass of the halos associated
with the QSOs, providing indications of QSO lifetimes and making
it possible to constrain QSO evolutionary models (Croom et al. 2005, da
$\hat{A}$ngela et al. 2008).

Redshift space distortions of the clustering pattern also contain
dynamical information on the QSO bias that are independent of any
assumption about underlying mass clustering. The clustering is affected
at small scales by the rms velocity dispersion of QSOs along the
line-of-sight (Fingers of god) and by dynamical infall of matter
into higher density regions, which causes a flattening of the clustering
pattern in the redshift direction. In addition to  these dynamical
distortions, geometrical distortions are introduced if an incorrect
cosmological model is used in order to convert the observed redshifts
into comoving distances. They therefore also constrain, more weakly, the
value of the cosmological density parameter, $\Omega_m$.

In the linear regime of clustering, dynamical infall is governed by the
parameter $\beta =\Omega_m^{0.6}/b$. (Peebles, 1980, Kaiser 1987,
Loveday et al. 1996, Matsubara \& Suto 1996, Matsubara \& Szalay 2001,
Peacock et al. 2001, Hoyle et al. 2002, Coil et al. 2005, Myers et al. 2006, 2007, Porciani \& Norberg 2006, da $\hat{A}$ngela et al. 2008, Ross et al. 2007). In recent years, measurements of QSO clustering (da $\hat{A}$ngela et al. 2008) yielded a $\beta_{QSO}(z=1.4)=0.60_{-0.11}^{+0.14}$ and $b_{QSO}(z=1.4)=1.5\pm0.2$ for a combined sample of QSOs from the
2dF QSO Redshift Survey (2QZ, Croom et al. 2004) and the 2dF-SDSS LRG
and QSO Survey (2SLAQ, Cannon et al. 2006). Ross et al. (2007) performed
similar measurements on the 2SLAQ Luminous Red Galaxies (LRGs)
clustering and found $\beta_{LRG}(z=0.55)=0.45_{-0.05}^{+0.05}$ and
$b_{LRG}(z=0.55)=1.66\pm0.35$.

In this paper we use QSOs from the 2QZ, 2SLAQ and SDSS (York et al. 2000) Data Release 5 (DR5; Adelman-McCarthy 2007) surveys and
LRGs from 2SLAQ and AAOmega first to  study the dependence of QSO-LRG
clustering amplitude on QSO luminosity. We also measure QSO-LRG redshift
distortions to estimate the  dynamical infall parameter, $\beta$.  These
surveys provide large numbers of QSOs and LRGs
with a range of luminosities at fixed redshifts. So our results, for
example on QSO bias, should be statistically improved over those from
QSO-QSO clustering. Moreover, we use photometric LRG samples, which
are significantly larger than the spectroscopic ones and measure QSO 2-D
correlation function amplitudes, using Limber's formula to convert to
3-D real-space measurements.

In section 2 we describe the QSO and LRG samples we use in our
measurements. In section 3 we present results from the 2-point
cross-correlation function $\omega (\theta)$, in Sections 4 and 5 we
show our results for the redshift-space cross-correlation function, $\xi
_s$, and the semi-projected cross-correlation function,
$w_p(\sigma)/\sigma$, respectively. Section 6 has our results for the
real-space cross-correlation function and in Section 7 we present our
$\xi (\sigma, \pi )$ results, model the redshift-space
distortions and estimate the QSO infall mass and bias. We also study, in
Section 8, the dependence of the redshift-space
cross-correlation function, the QSO bias and the mass of the QSO-LRG
Dark Matter Haloes ($M_{DMH}$) on QSO luminosities at fixed redshifts.
Finally, in Section 9 we present our conclusions.

\section{Data}

\subsection{Spectroscopic data}

Our spectroscopic LRG sample is taken from the 2SLAQ and consists of 8,656 LRGs within $0.35\leq z\leq 0.75$ (that is the $`$Gold Sample' of Ross et al. (2007), see their Fig. 1 for the LRG distribution). 5,995 LRGs are in the Northern strip (sectors a, b, c, d, e, from Fig. 1 of Ross et al. 2007) and 2,661 LRGs in the Southern strip (sector s).

Our QSOs are taken from three spectroscopic samples; the NGC of 2QZ, the NGC+SGC of 2SLAQ and the Data Release 5 (DR5) of SDSS. The QSO redshift range mainly used in our analysis is the same one as for the 2SLAQ LRGs ($0.35\leq z\leq 0.75$). The 2SLAQ QSOs have the same distribution on the sky as the spectroscopic 2SLAQ LRGs (see Fig. 2 of da $\hat{A}$ngela et al. 2008). The 2QZ and SDSS QSOs cover only the NGC of the 2SLAQ LRGs. The brightest of our QSO samples is from SDSS, which consists of QSOs with $i_{AB}< 19.1$. Our 2QZ sample consists of QSOs with $18.25\leq b_J\leq 20.85$ and the 2SLAQ sample of QSOs with $18.0\leq g\leq 21.85$. After matching the QSO and spectroscopic 2SLAQ LRG areas we get the numbers shown in Table \ref{Table:spec_qso}.

\begin{table}
\caption{The numbers of spectroscopic QSOs and spectroscopic 2SLAQ LRGs.}
\centering
\setlength{\tabcolsep}{1.5mm}
\begin{tabular}{lcc}
       \hline
$$ & 2SLAQ area ($0.35\leq z\leq 0.75$)\\
       \hline 
$$ & QSOs & LRGs \\
       \hline
2SLAQ & $699$ & $8,656$   \\
       \hline 
2QZ & $307$ & $5,995$ \\
       \hline
SDSS & $218$ & $5,995$  \\
       \hline
\label{Table:spec_qso}
\end{tabular}
\end{table}

\begin{table}
\caption{The numbers of spectroscopic QSOs and photometric LRGs.}
\centering
\setlength{\tabcolsep}{0.5mm}
\begin{tabular}{lcccc}
       \hline
$$ & \multicolumn{2}{c}{2SLAQ area ($0.35\leq z\leq 0.75$)} &\multicolumn{2}{c}{AAOmega area ($0.45\leq z\leq 0.90$)} \\
       \hline
$$ & QSOs & LRGs & QSOs & LRGs \\
       \hline
2SLAQ & $503$ & $19,300$ & $786$ & $23,836$  \\
       \hline 
2QZ & $1,048$ & $32,188$ & $1,265$ & $40,060$ \\
       \hline
SDSS & $7,395$ & $468,416$ & $7,083$ & $571,676$ \\
       \hline
\label{Table:photo_qso}
\end{tabular}
\end{table}

\subsection{Photometric data}

For measuring the 2-point angular cross-correlation function, $w(\theta)$, we can also use larger, photometric LRG samples from the 2SLAQ and the AAOmega surveys. For the AAOmega LRGs we have followed the selection criteria of Ross et al. (2007). The numbers for these LRG data sets and the new matched QSO samples are shown in Table \ref{Table:photo_qso}. We should mention here that in the case of photometric 2SLAQ and AAOmega LRGs with the 2SLAQ QSOs we did not perfectly match the two areas. The photometric 2SLAQ LRG area is a rectangle ($137.5^{\circ}\leq \rm{ra}\leq 230.0^{\circ}$, $-1.25^{\circ}\leq \rm{dec}\leq 1.0^{\circ}$) which covers the whole NGC of 2SLAQ, including the gaps in between the sectors a, b, c, d, e, whereas the 2SLAQ QSOs have the distribution mentioned in the previous Section. This slight mis-match does not affect our results, but is the reason that, in Table \ref{Table:photo_qso}, photometric 2SLAQ LRGs matched with 2SLAQ QSOs appear more numerous than the spectroscopic 2SLAQ LRGs matched with 2SLAQ QSOs in Table \ref{Table:spec_qso} (19,300 vs. 8,656). Furthermore, the 2SLAQ QSO set, in the 2SLAQ photometric area, is smaller than that in the 2SLAQ spectroscopic area (503 vs 699) because in the photometric case we do not use the SGC part of 2SLAQ (sector s). 

\begin{figure*}
\begin{center}
\centerline{\epsfxsize = 9.0cm
\epsfbox{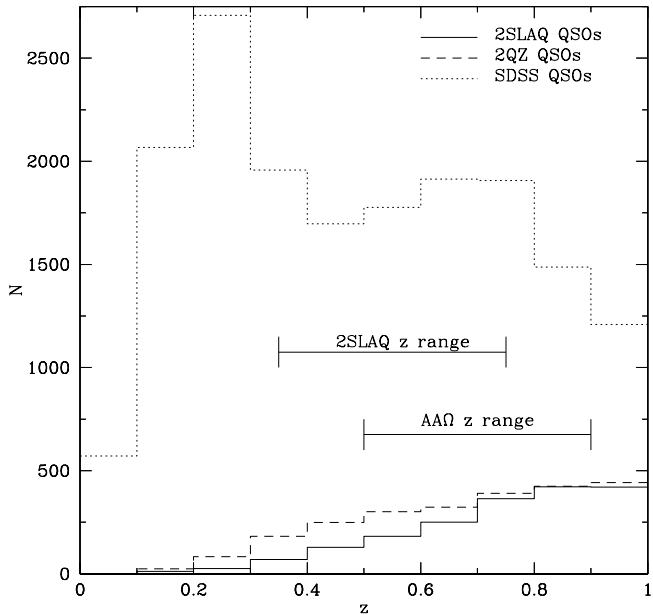}}
\caption{The solid line shows N(z) for the 2SLAQ sample, the dashed
 line for the 2QZ and the dotted line for the SDSS sample. The long
 dashed line and the dashed-dotted line show the redshift range we use for
 our measurements, when we cross-correlate our QSO samples with the
 2SLAQ and AAOmega LRGs, respectively. The 2SLAQ and 2QZ distributions
 are, as expected, very similar. The SDSS distribution is flat, in the 2SLAQ and AAOmega redshift ranges we are interested in.}
\label{fig:n_z_qsos}
\end{center}
\end{figure*}

The redshift distribution of all the QSO samples, in a redshift range of $0.0<z\leq 1.0$ is shown in Fig. \ref{fig:n_z_qsos}. The 2SLAQ and 2QZ distributions are, as expected, very similar. The SDSS distribution is roughly flat, in the redshift range we are interested in, although we note a peak at $z\simeq0.2$ which is outside our region of interest.

\section{QSO-LRG angular cross-correlation function}

\subsection{Cross-correlation and error estimators}

In this section we first estimate the 2SLAQ QSO$-$2SLAQ LRG and 2QZ
 QSO$-$2SLAQ LRG angular cross-correlation functions $w(\theta )$ using our
 spectroscopic samples, both for QSOs and LRGs. We then use our photometric 2SLAQ and AAOmega
 LRG samples and cross-correlate them with 2SLAQ, 2QZ and SDSS (DR5)
 QSOs. The estimator we use, in all cases, is 

\begin{equation}
w(\theta )=\frac{DD(\theta )}{DR(\theta )}\frac{N_{rd}}{N_{LRG}}-1
\label{eqn:wtheta}
\end{equation}
where $N_{rd}$ is the number of points in our random catalogue, $N_{LRG}$ is the number of LRGs, $DD(\theta)$ are the data-data pairs, i.e. QSO-LRG pairs and $DR(\theta)$ are the QSO-random point pairs counted at angular separation $\theta$. 

 Our spectroscopic LRG random catalogue is the same as that described by Ross et al. (2007) (with $20\times$ more randoms than LRGs) whereas the photometric 2SLAQ and AAOmega LRG random catalogues contain $\simeq11\times$ more randoms than LRGs. Coverage completeness and spectroscopic completeness are taken into account in the construction of all random catalogues. 

The error estimator we use throughout this paper is the so-called Field-to-Field estimator. The accuracy of the errors on our measurements plays an important role, particularly when we use our results to model the redshift-space distortions in Section 8. The reason for not using a different error estimator, such as the Poisson estimator ($\sigma(\theta)=\frac{\sqrt{DD(\theta)}}{DR(\theta)}$) is that it becomes increasingly inaccurate at larger scales, as the QSO-LRG pairs become less independent. In order to calculate the Field-to-Field errors we have divided the 2SLAQ area into 4 approximately equal areas and then measure the cross-correlation functions in each one of these areas. The Field-to-Field error is then given by the following expression (Myers et al. 2005)

\begin{equation}
\sigma_\omega^2(\theta) =
\frac{1}{N-1}\sum_{L=1}^{N}\frac{DR_L(\theta)}{DR(\theta)}[\omega_L(\theta)-\omega(\theta)]^2
\end {equation}
where $DR_L(\theta)$ is the data-random pairs in the subarea,
 $DR(\theta)$ is the overall number of data-random pairs, $\omega_L(\theta)$ is
 the correlation function measured in the subarea and $\omega(\theta)$ is
 the overall correlation function.

\subsection{$w(\theta )$ results from the redshift samples and correction for fibre collision}

Fig. \ref{fig:2D_2slaq} shows our angular correlation results when we cross-correlate 2SLAQ QSOs with spectroscopic 2SLAQ LRGs. Filled circles show the results when we use our whole 2SLAQ QSO sample and cross-correlate it with our sample 8 2SLAQ LRGs. Open circles show the results from 2SLAQ QSOs in the redshift range of $1.0\leq z\leq 2.2$ cross-correlated with the same LRG sample. Finally, triangles show the results when we use 2SLAQ QSOs in the redshift range of $0.35\leq z\leq 0.75$. The fibre collision problem affects our results as can be seen from the anti-correlation we find, in all cases, between our QSOs and LRGs on small scales. The results are very similar when we use 2SLAQ QSOs in the whole redshift range and within $1.0\leq z\leq2.2$. The anti-correlation amplitude is increased when we use our low redshift QSO sample, which also shows an increased fibre effect, but its statistical significance is smaller as the sample is much smaller than the other two. Now, we shall use these results in order to correct our 2-D and 3-D results, for fibre collisions. 

We shall base our correction on the Hawkins et al. (2003) expression, i.e.

\begin{equation}
w_f=\frac{1+w_p}{1+w_z}
\end{equation}
$w_p$ is the cross-correlation results when our samples come from the input catalogue; in our QSO-LRG case $w_p=0$ if there is no redshift overlap between QSOs and LRGs and assuming no lensing. $w_z$ represents the cross-correlation results from the 2SLAQ catalogue and these are the results shown in Fig. \ref{fig:2D_2slaq}. The correction should therefore be

\begin{equation}
w_f=\frac{1}{1+w_z}
\end{equation}

We determine $w_z$ based on the measurements shown in Fig. \ref{fig:2D_2slaq}. As already noted, on small scales ($\leq2'$), we see the anti-correlation caused by the fibre effect. We also note a small positive bump on scales up to $\sim80'$. This bump is expected to be caused by physical QSO-LRG cross-correlations for the $0.35\leq z\leq 0.75$ range (triangles) and also to a lesser extent in the unrestricted redshift range (closed circles). The cause of small positive excess seen at the $1.0\leq z\leq 2.2$ could be due to the fibre collision effect. If we include this small positive bump in the fibre correction, our $\xi (s)$ cross-correlation measurements do not change; therefore this bump will be ignored when we present our 3-D results in the next Section. Nevertheless, the feature's effect is larger for the 2SLAQ QSO-2SLAQ (photometric) LRG $w(\theta )$ measurements which will be presented later (Fig. \ref{fig:2SLAQ_LRG_3sur}) and it will be taken into account there.

Finally, Fig. \ref{fig:2D_2qz} shows our results when we cross-correlate 2QZ QSOs with spectroscopic 2SLAQ LRGs. No fibre collision effect is expected in this case since 2QZ QSOs had higher priority than 2dFGRS galaxies for spectroscopic observations. Thus there is no issue of fibre incompleteness (see also, e.g., Myers et al. 2003). The reason for performing these measurements is to see if we can detect any signal in the case of the overlapped redshift range (triangles). Although the results when we use 2QZ QSOs in $1.0\leq z\leq 2.2$ range (open circles) seem to give a null average signal as expected, in the case of the common QSO-LRG redshift range (triangles) the samples are small and the results appear too noisy to draw any statistically significant conclusions.

\subsection{Results from the photometric samples}

Before we proceed to measure the redshift-space cross-correlation function, we shall repeat our $w(\theta )$ measurements, cross-correlating our spectroscopic QSO samples from SDSS, 2QZ and 2SLAQ with the photometric LRG samples, which were presented in Section 2.2 and Table \ref{Table:photo_qso}. The purpose of these measurements is to see how the LRGs correlate with bright and faint QSOs and then use Limber's formula to convert these 2-D measurements into 3-D real-space measurements and compare them with our results from the spectroscopic samples (Section 4). 

\begin{figure*}
\begin{center}
\centerline{\epsfxsize = 9.0cm
\epsfbox{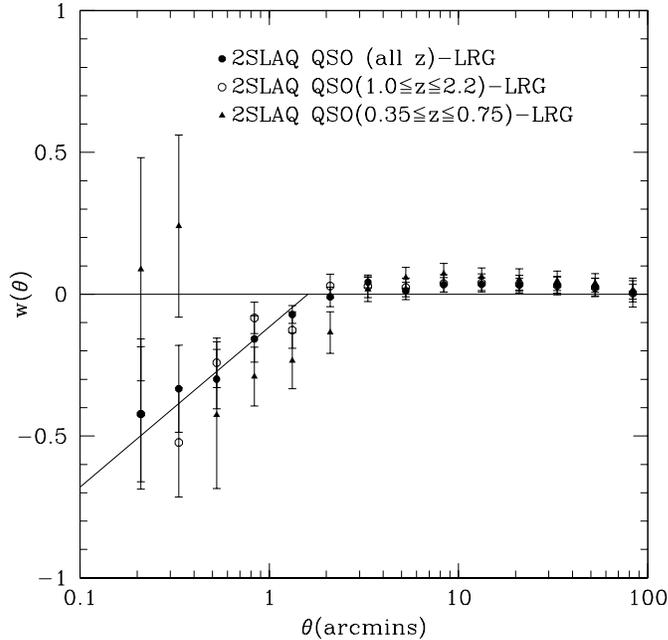}}
\caption{2SLAQ QSO-2SLAQ (spectroscopic) LRG cross-correlation results. Filled circles show the results when we use our entire (9,044) 2SLAQ QSO sample and cross-correlate it with our spectroscopic 2SLAQ LRGs (9,856). Open circles show the results when we use (6,002) 2SLAQ QSOs in the redshift range of $1.0\leq z\leq 2.2$ and finally, triangles show the results when we use (699) 2SLAQ QSOs with a redshift range of $0.35\leq z\leq 0.75$. The fibre collision effect is seen in the results as we get an anti-correlation between our QSOs and LRGs. The solid line shows our best fit to the results (filled circles).}
\label{fig:2D_2slaq}
\end{center}
\end{figure*}

The results using 2SLAQ photometric LRGs are shown in Fig. \ref{fig:2SLAQ_LRG_3sur} (filled circles are corrected for fibre collisions, from $0.1'<\theta<100'$, i.e. including the bump in Fig. \ref{fig:2D_2slaq}) and the results using the AAOmega photometric LRGs are shown in Fig. \ref{fig:AAO_LRG_3sur}. We should note that neither the AAOmega nor the 2SLAQ results have been corrected for stellar contamination of the LRGs ($\simeq15\%$ in AAOmega and $\approx5\%$ in 2SLAQ). In both cases, triangles show the results using SDSS QSOs, filled circles using 2SLAQ QSOs and open circles using 2QZ QSOs. The $\theta _0$ and $r_0$ values from the fits (in the range 0.1$'$-100$'$ ) to these measurements are also shown. Both results show a slightly steeper slope for the brightest QSO sample (SDSS, $\gamma =-0.8\pm0.1$) compared to the other QSO samples.

\begin{figure*}
\begin{center}
\centerline{\epsfxsize = 9.0cm
\epsfbox{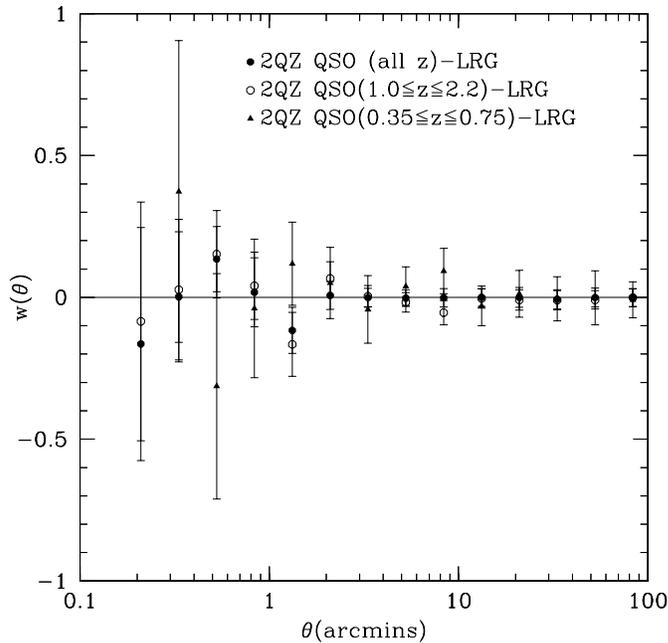}}
\caption{2QZ QSO-2SLAQ LRG cross-correlation results. Filled circles
 show the results when we use our entire (2,854) 2QZ QSO sample ($`$11' quality, e.g., Croom et al. 2004) and cross-correlate it with our spectroscopic 2SLAQ LRGs (NGC, 5,995 LRGs). Open circles show the results when we use (1,699) 2QZ QSOs in the redshift range of $1.0\leq z\leq 2.2$ cross-correlated with the same LRG sample. Finally, triangles show the results when we use (307) 2QZ QSOs with redshift range $0.35\leq z\leq 0.75$.}
\label{fig:2D_2qz}
\end{center}
\end{figure*}

\begin{figure*}
\begin{minipage}[t]{1.0\linewidth}
\centerline{\epsfxsize = 10.0cm
\epsfbox{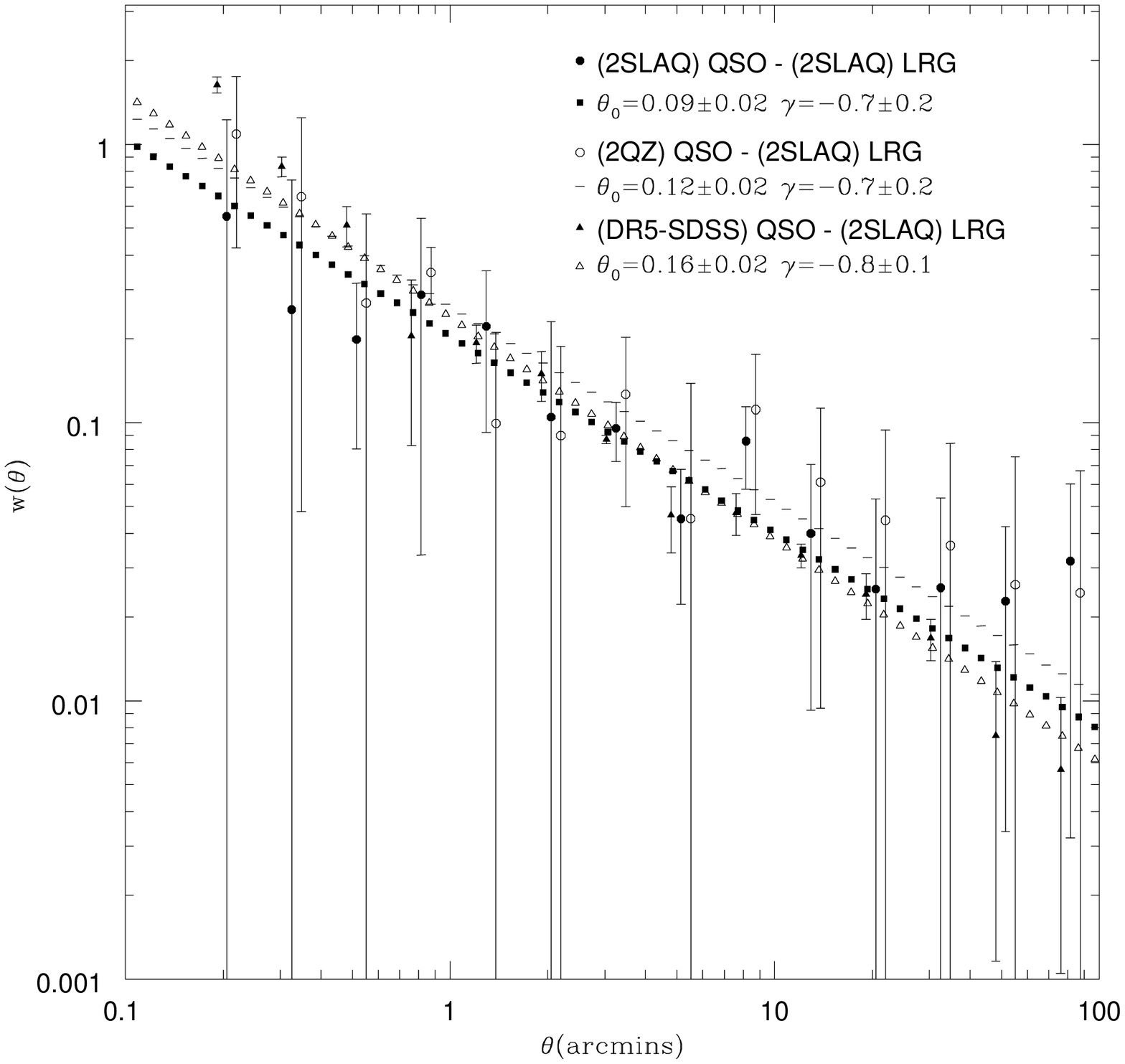}}
\caption{Triangles show the results from SDSS QSOs, filled circles from 2SLAQ QSOs (corrected for fibre collisions) and open circles from 2QZ QSOs cross-correlated with photometric 2SLAQ LRGs. We have also plotted the fits to these measurements. We note that the slope is steeper for the brightest SDSS QSO sample comparing to the fainter 2QZ and 2SLAQ QSO samples.}
\label{fig:2SLAQ_LRG_3sur}
\end{minipage}
\begin{minipage}[b]{1.0\linewidth}
\centerline{\epsfxsize = 10.0cm
\epsfbox{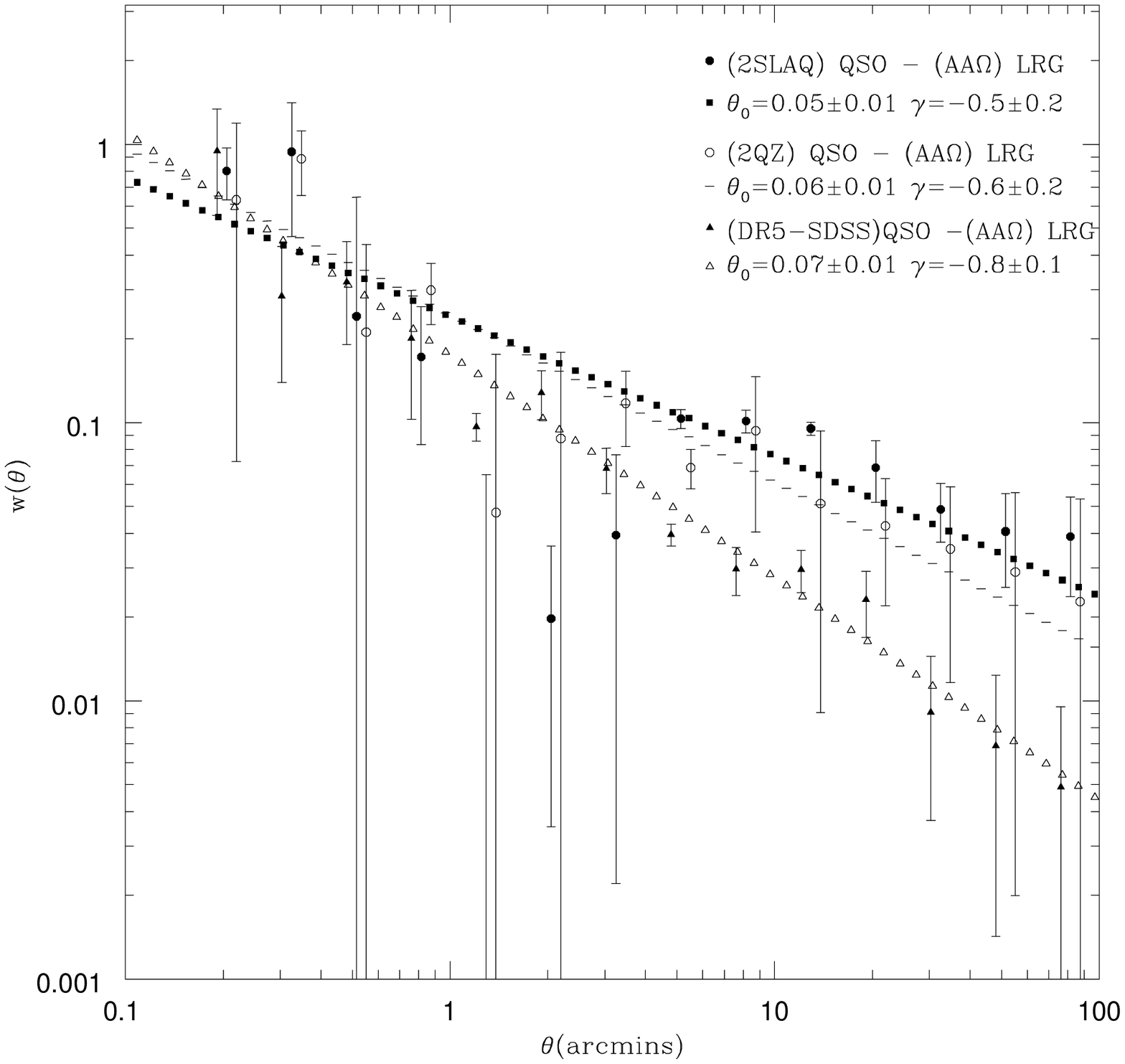}}
\caption{Triangles show the results from SDSS QSOs, filled circles from 2SLAQ QSOs and open circles from 2QZ QSOs cross-correlated with photometric AAOmega LRGs. We have also plotted the fits to these measurements. Once more, we note that the slope is steeper for the brightest SDSS QSO sample comparing to the fainter 2QZ and 2SLAQ QSO samples.}
\label{fig:AAO_LRG_3sur}
\end{minipage}
\end{figure*}

Comparing our $w( \theta)$ results from 2QZ QSO-spectroscopic LRGs (triangles of Fig. \ref{fig:2D_2qz}) with those from 2QZ QSO-photometric LRGs (open circles of Fig. \ref{fig:2SLAQ_LRG_3sur}) we note that the latter shows a positive signal whereas the former shows no signal, at least on small scales, and appears to be noisy. This is probably because the 2QZ QSO sample is much larger ($\approx3\times$) in the photometric LRG area (Table \ref{Table:photo_qso}) than it is in the spectroscopic LRG area (Table \ref{Table:spec_qso}).  

\section{3-D cross-correlation functions, $\xi\MakeLowercase{(s)}$ and $\xi \MakeLowercase{(r)}$}

\begin{figure*}
\begin{center}
\centerline{\epsfxsize = 9.0cm
\epsfbox{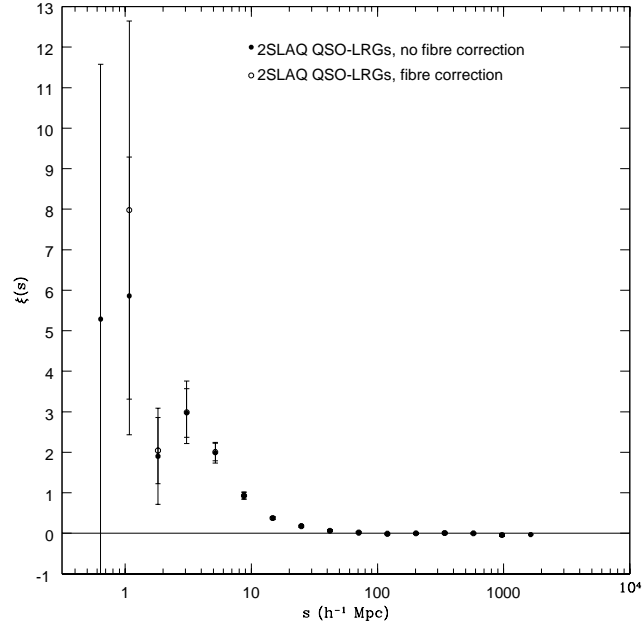}}
\caption{2SLAQ QSO-2SLAQ (spectroscopic) LRG redshift-space
 cross-correlation results. Filled circles show the results when the fibre collision effect is not taken into account and open circles show the results
 when the fibre collision is taken into consideration. As we can see the effect is bigger on small scales (i.e. $s\leq 3$h$^{-1}$Mpc) than it is on larger scales.}
\label{fig:xis_2slaq}
\end{center}
\end{figure*}

In this Section we present our results for the QSO-LRG redshift-space cross-correlation function, $\xi(s)$, for different QSO and LRG samples. Fig. \ref{fig:xis_2slaq} shows the results using our 2SLAQ QSO-2SLAQ (spectroscopic) LRG samples. Our 2SLAQ QSO sample consists of QSOs with the same average redshift as the 2SLAQ LRGs, i.e. $0.35\leq z\leq 0.75$. Filled circles show the results when the fibre collision effect is not taken into account and open circles show the results when the fibre collision is taken into consideration, as described in the previous Section. As we can see the effect is bigger on small scales (i.e. $s\leq 3$h$^{-1}$Mpc) than it is on larger scales.

\begin{figure*}
\begin{center}
\centerline{\epsfxsize = 9.0cm
\epsfbox{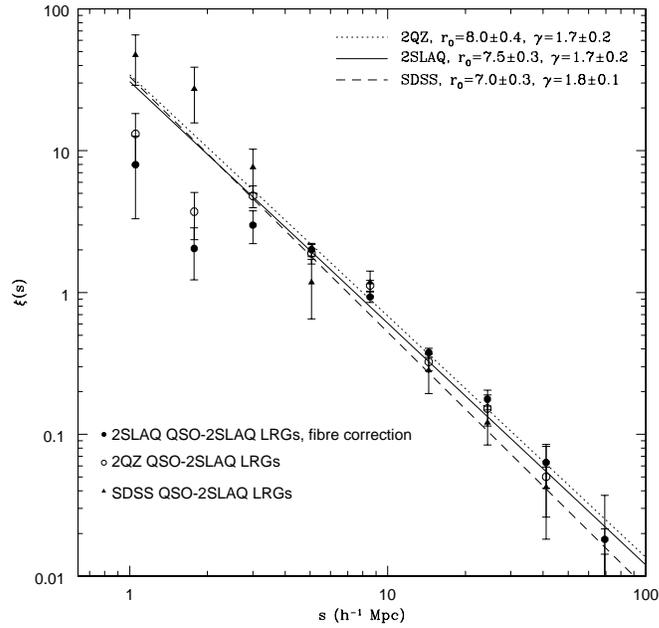}}
\caption{QSO-2SLAQ LRG redshift-space cross-correlation results.  Filled
 circles show the results when using 2SLAQ QSOs, open circles using 2QZ QSOs
 and triangles using SDSS DR5 QSOs ($0.35\leq z\leq 0.75$, in all
 cases). All measurements have been made with spectroscopic 2SLAQ LRGs. The lines show the $\xi (r)$ fits from the photometric samples, which appear to be in agreement with the spectroscopic results.}
\label{fig:xis_all1}
\end{center}
\end{figure*}

We also repeat this redshift-space cross-correlation measurement using the 2QZ and the SDSS QSO samples, both within a redshift range of $0.35\leq z\leq 0.75$, and spectroscopic 2SLAQ LRGs. As already mentioned, the 2QZ survey has a brighter magnitude limit than 2SLAQ, i.e. $b_J=20.85$ instead of $b_J=21.85$. The brightest of the three samples is that from SDSS ($i_{AB}<19.1$). Our results are shown in Fig. \ref{fig:xis_all1}. On small scales, $\leq 5$h$^{-1}$Mpc we can see the suppression of $\xi (s)$, due to the non-linear redshift-space distortions (small-scale peculiar velocities, $\langle w_z^2\rangle ^{1/2}$).  The effect is less for SDSS, possibly because SDSS uses narrow lines to determine redshifts for QSOs at low redshift and [OIII] 5007 is seen for the whole redshift range in question for SDSS spectra. 2SLAQ/2QZ redshifts are purely template fits and so will instead be driven by the broad lines. On larger scales the results are in good statistical agreement, regardless of the brightness of the QSO sample. This agreement is also confirmed by the fits to these measurements (Table \ref{table:spectroscopic}). Since the results are affected by the Finger of god effect on small scales, the fits are applied on $5-25$h$^{-1}$Mpc scales. The agreement between the $\xi (s)$ results, suggests that QSO bias is also independent of QSO luminosity since the QSOs span a range of more than 2 magnitudes at fixed redshift.

\begin{figure*}
\begin{center}
\centerline{\epsfxsize = 9cm
\epsfbox{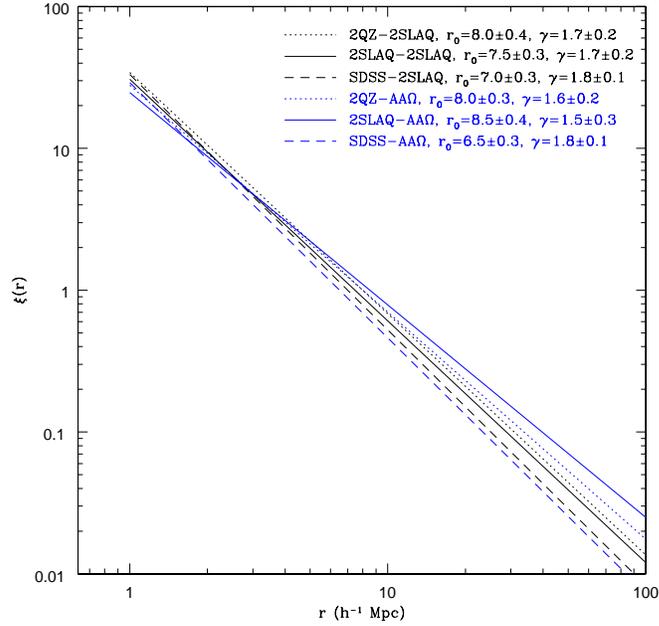}}
\caption{$\xi (r)$ fits via Limber's formula following Phillipps
 et al. 1977, of our $w(\theta )$ measurements from spectroscopic QSO
 samples with photometric 2SLAQ and AAOmega LRG samples.}
\label{fig:Limber_all}
\end{center}
\end{figure*}

To further check our observations, we use Limber's formula (Limber 1953, Rubin 1954) and following Phillipps et al. (1978) we calculate the real-space cross-correlation function, $\xi (r)$ from our previous (Section 3) $w(\theta )$ measurements from the photometric LRG samples. The fits appear in Fig. \ref{fig:Limber_all} and in Table \ref{table:photometric}. The black solid line ($r_0=7.5\pm0.3, \gamma=1.7\pm0.2$) is the fit using the 2SLAQ QSO-2SLAQ LRG sample, the dotted line ($r_0=8.0\pm0.4, \gamma=1.7\pm0.2$) using the 2QZ QSO-2SLAQ LRG sample and the dashed line ($r_0=7.0\pm0.3, \gamma=1.8\pm0.1$) using the SDSS QSO-2SLAQ LRG sample. The blue lines are the fits using our QSO samples with AAOmega LRGs. Once again, we note that the results are approximately independent of the luminosity of the QSO sample.

In Fig. \ref{fig:xis_all1} we have plotted the fits from the QSO-2SLAQ (photometric) LRGs with the results from the spectroscopic 2SLAQ LRGs already discussed. This is a comparison between $\xi (r)$ (photometric) and $\xi (s)$ (spectroscopic) results. As we can see from Fig. \ref{fig:xis_all1} the agreement is not good at small scales ($<5$h$^{-1}$Mpc) but this is due to non-linear redshift-space distortions (Finger of god) that affect $\xi (s)$ but not the $\xi (r)$ measurements on small scales. A fairer comparison can be made on larger scales. Taking into account the Kaiser boost ($\approx1.25$ for $\beta =0.35$, as we shall see next) we compare the fits (Tables \ref{table:photometric} and \ref{table:spectroscopic}) to the QSO-2SLAQ (photometric) LRGs and QSO-2SLAQ (spectroscopic) LRGs, respectively. We see that they are in reasonable agreement.

\begin{figure*}
\begin{center}
\centerline{\epsfxsize = 9.0cm
\epsfbox{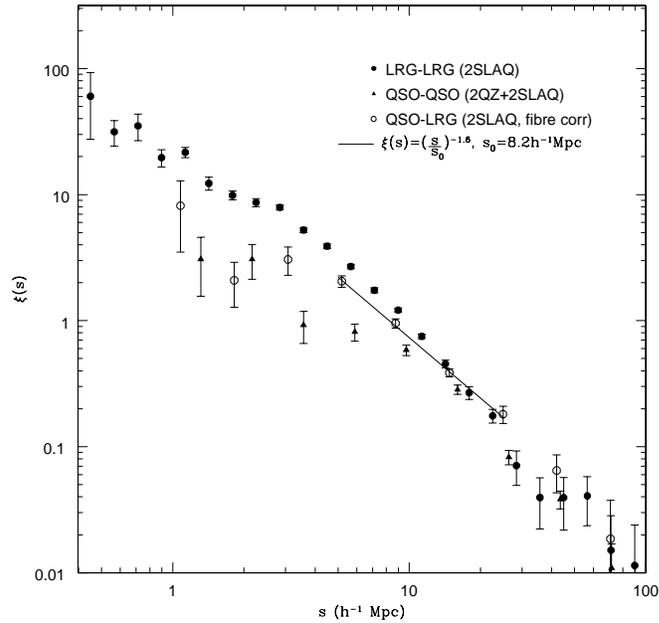}}
\caption{Comparison of $\xi (s)$ measurements for QSO-LRGs with
 QSO-QSO (da $\hat{A}$ngela et al. 2008) and LRG-LRG (Ross et al. 2007)
 measurements. Triangles are the 2QZ+2SLAQ QSO $\xi(s)$ results ($0.3<z<2.2$), filled
 circles show the 2SLAQ LRG $\xi (s)$ results and open circles show the
 results for the 2SLAQ QSO-2SLAQ LRG redshift-space cross-correlation.
 The solid line
 shows our $\chi^2$ fit to the data from $5-25$h$^{-1}$Mpc, which gives
 $s_0=8.2\pm0.1$h$^{-1}$Mpc and $\gamma=1.6_{-0.1}^{+0.2}$.}
\label{fig:xis_all2}
\end{center}
\end{figure*}

Finally, in Fig. \ref{fig:xis_all2} we compare our redshift-space measurements for QSO-LRGs with QSO-QSO (da $\hat{A}$ngela et al. 2008) and LRG-LRG (Ross et al. 2007) measurements. Triangles show the 2QZ+2SLAQ QSO $\xi(s)$ results, filled circles show the 2SLAQ LRG $\xi (s)$ results and open circles show our results for the 2SLAQ QSO-2SLAQ (spectroscopic) LRG redshift-space cross-correlation. The fact that QSO-QSO and QSO-LRG results appear flatter than the LRG-LRG ones, on small scales, may be explained by the intrinsic dispersion and the redshift errors, which are higher for broad emission line QSOs than for LRGs ($\approx650$kms$^{-1}$ vs. $\approx300$kms$^{-1}$). This would affect mostly the QSO-QSO results and the least the LRG-LRG results, as observed. The QSO-LRG measurements should then lie in between the QSO-QSO and LRG-LRG measurements (as they do). On larger scales ($\geq 5$h$^{-1}$Mpc) the QSO-QSO correlation amplitude appears lower than the LRG-LRG one. This implies that the QSO bias is smaller than the LRG bias. Assuming the model $\xi _{mm}=\frac{\xi _{QL}}{b_Qb_L}$ (see Section 7.2), the agreement between the QSO-LRG amplitude and the LRG-LRG amplitude would imply that $b_Q\approx b_L$. So the overall conclusion is that $b_Q\lesssim b_L$. These issues will be further discussed in Section 8.

\begin{figure*}
\begin{center}
\centerline{\epsfxsize = 9.0cm
\epsfbox{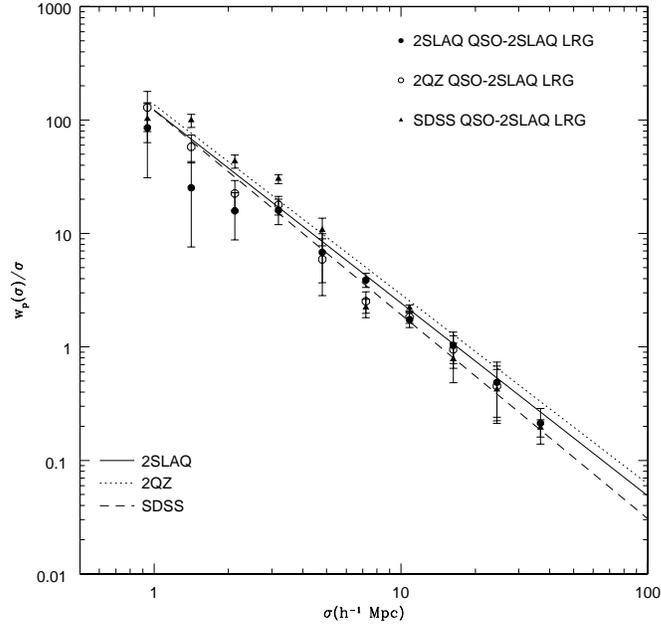}}
\caption{The semi-projected cross-correlation function results for the 2SLAQ QSOs-2SLAQ (spectroscopic) LRGs (filled circles), the 2QZ QSOs-2SLAQ (spectroscopic) LRGs (open circles) and SDSS QSOs-2SLAQ (spectroscopic) LRGs (triangles). We have also plotted the fits from the $w(\theta )$ measurements of the photometric 2SLAQ LRG sample, using Limber's formula.}
\label{fig:projected1}
\end{center}
\end{figure*}

\begin{figure*}
\begin{center}
\centerline{\epsfxsize = 9.0cm
\epsfbox{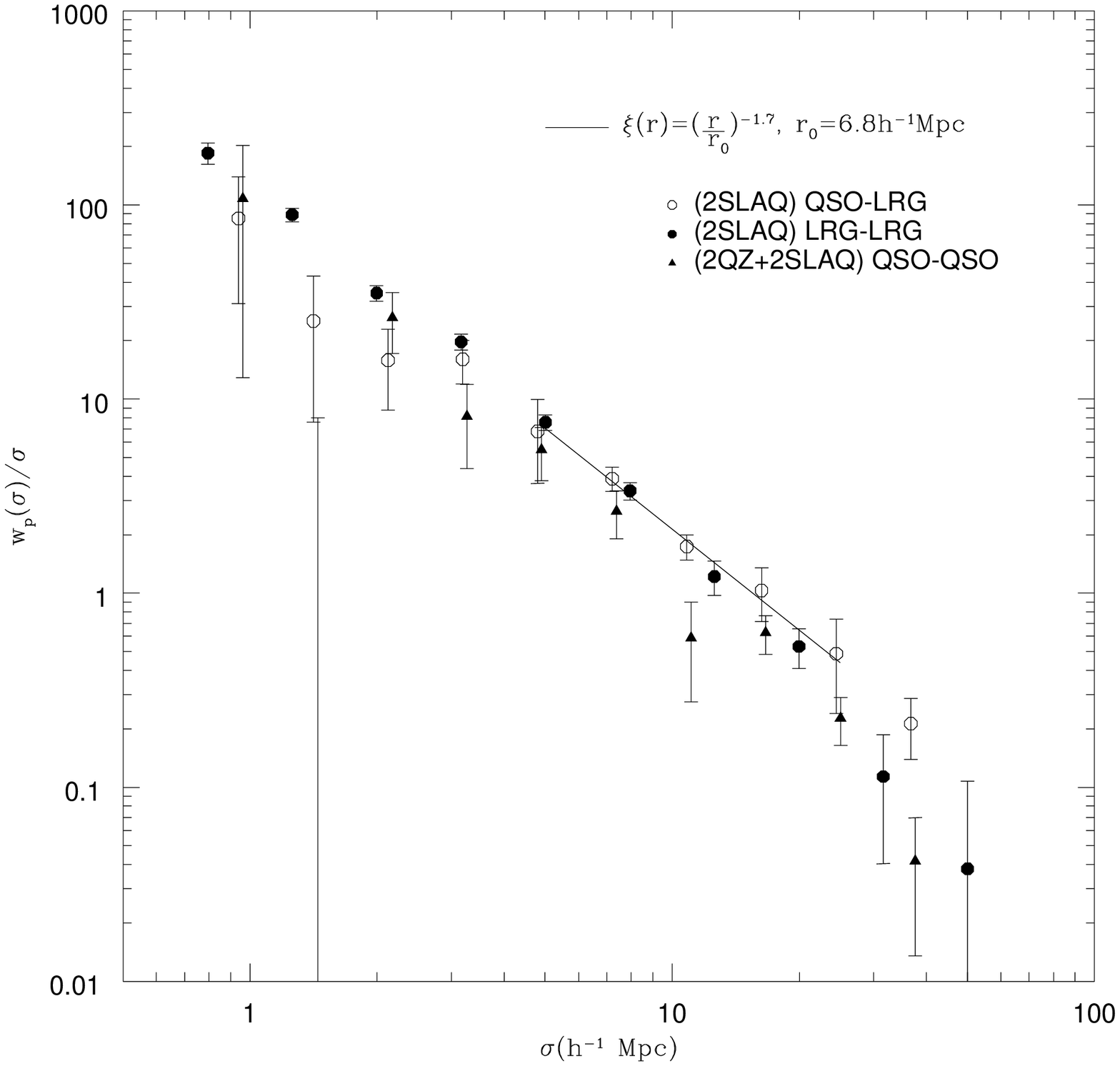}}
\caption{The semi-projected correlation function results for the (2QZ+2SLAQ) QSO (open circles) and the 2SLAQ LRG-LRG (triangles) from da $\hat{A}$ngela et al. (2008) and Ross et al. (2007), respectively. The solid line shows our $\chi^2$ fit to the data from $5-25$h$^{-1}$Mpc, which gives $r_0=6.8_{-0.3}^{+0.1}$h$^{-1}$Mpc and $\gamma=1.7_{-0.3}^{+0.2}$.}
\label{fig:projected2}
\end{center}
\end{figure*}

\section{The semi-projected cross-correlation function}

If $s_1$ and $s_2$ are the distances of two objects 1, 2, measured in
 redshift-space, and $\theta$ the angular separation between them, then
 $\sigma$ and $\pi$ are defined as

\begin{equation}
\pi=(s_2-s_1), $ along the line-of-sight$
\end{equation}

\begin{equation}
\sigma=\frac{(s_2+s_1)}{2}\theta , $ across the line-of-sight$
\end{equation}
The effects of redshift distortion appear only in the radial component,
 $\pi$; integrating along the $\pi$ direction, we calculate what
 is called the projected correlation function, $w_p(\sigma)$

\begin{equation}
w_p(\sigma)=2\int_0^\infty \xi(\sigma,\pi)d\pi
\end{equation}
In our case we take the upper limit of the integration to be equal to
 $\pi_{max}=70$h$^{-1}$Mpc. This limit has been chosen to minimise the effect of the small-scale peculiar velocities and redshift errors (da $\hat{A}$ngela et al. 2008). If we include very large scales, the signal will become dominated by noise and; on
 the other hand, if we take our measurements on very small scales then
 the amplitude will be underestimated. Now, since $w_p(\sigma)$ describes
 the real-space clustering, the last equation can be written in terms
 of the real-space correlation function, $\xi(r)$, (Davis \& Peebles,
 1983), i.e.

\begin{equation}
w_p(\sigma)=2\int_\sigma^{\pi_{max}}\frac{r\xi (r)}{\sqrt(r^2-\sigma
 ^2)}dr
\end{equation}
Calculating the projected cross-correlation function $w_p(\sigma)$
 will help us estimate the real-space cross-correlation function,
 $\xi(r)$. The $r_0$ from the $w_p(\sigma)/\sigma$ fits is estimated using the following equation:

\begin{equation}
\frac{w_p(\sigma )}{\sigma}=\left(\frac{r_0}{\sigma}\right)^\gamma \frac{\Gamma
 (\frac{1}{2})\Gamma (\frac{\gamma-1}{2})}{\Gamma(\frac{\gamma}{2})}
\label{eqn:projected_ro}
\end{equation}
where $\Gamma (x)$ is the Gamma function.

Fig. \ref {fig:projected1} shows our $w_p(\sigma)/\sigma$ measurements from the different QSO samples cross-correlated with the spectroscopic 2SLAQ LRGs. Filled circles show the results using 2SLAQ QSOs-2SLAQ LRGs, open circles using 2QZ QSOs-2SLAQ LRGs and triangles SDSS QSOs-2SLAQ LRGs. We see that the semi-projected cross-correlation function confirms our results in redshift-space, i.e. the measurements are in agreement regardless of the luminosity of the QSO sample. This can also be confirmed by the fits to these measurements shown in Table \ref{fig:table_wp}. As for the $\xi (s)$ case, we also include the fits from the $w(\theta )$ measurements of the photometric 2SLAQ LRG sample, using Limber's formula. The photometric fits are, again, in good agreement with the spectroscopic measurements, further supporting the idea that the cross-clustering is independent of QSO luminosity. 

Fig. \ref {fig:projected2} shows the $w_p(\sigma)/\sigma$ results for the 2SLAQ QSO-LRGs (filled circles). As in the previous Section, we have also included the semi-projected correlation function results for the (2QZ+2SLAQ) QSO (open circles) and the 2SLAQ LRG-LRG (triangles) from da $\hat{A}$ngela et al. (2008) and Ross et al. (2007), respectively. We note that, at small scales ($\leq 3$h$^{-1}$Mpc), although the results are noisier than for the $\xi (s)$ measurements, QSO-LRG and QSO-QSO measurements have a slightly smaller amplitude than the LRG-LRG one but to a much lesser degree than in the $\xi (s)$ measurements. This confirms our previous interpretation that the amplitude difference in the redshift-space measurements at small scales is due to the QSO redshift errors, an effect which does not affect the $w_p(\sigma)/\sigma$ measurements. On larger scales, $w_p(\sigma)/\sigma$ measurements confirm our previous observations for lower QSO-QSO amplitude comparing with the QSO-LRG and the LRG-LRG amplitude. Finally, the solid line shows our $\chi^2$ fit to the QSO-LRG data from $5-25$h$^{-1}$Mpc (for consistency reasons with the $\xi (s)$ fits), which gives $r_0=6.8_{-0.3}^{+0.1}$h$^{-1}$Mpc and $\gamma=1.7_{-0.3}^{+0.2}$. This is similar to the (2SLAQ) LRG-LRG auto-correlation amplitude ($r_0=7.45\pm0.35$h$^{-1}$Mpc) and both are higher than the (2QZ+2SLAQ) QSO-QSO amplitude ($r_0\simeq5.0$h$^{-1}$Mpc) at $z=1.4$.

\section{The real-space cross-correlation function}

Using the results from the projected cross-correlation function, described in the previous Section, and following Saunders et al. 1992, we can calculate the real-space cross-correlation function, $\xi (r)$, as follows:

\begin{equation}
\xi (r)=-\frac{1}{\pi}\int_r^\infty\frac{d\omega
 (\sigma)/d\sigma}{\sqrt{(\sigma ^2-r^2)}}d\sigma
\end{equation}
and assuming a step function for $w_p(\sigma)=w_i$ we finally get,

\begin{equation}
\xi(\sigma _i)=-\frac{1}{\pi}\sum_{j\geq i}\frac{\omega _{j+1}-\omega
 _j}{\sigma _{j+1}- \sigma _j}ln{(\frac{\sigma _{j+1}+\sqrt{\sigma_
 {j+1}^2-\sigma _i^2}}{\sigma _j+\sqrt{\sigma_ j^2-\sigma _i^2}})}
\end{equation}
for $r=\sigma _i$. 

\begin{figure}
\begin{center}
\centerline{\epsfxsize = 9.0cm
\epsfbox{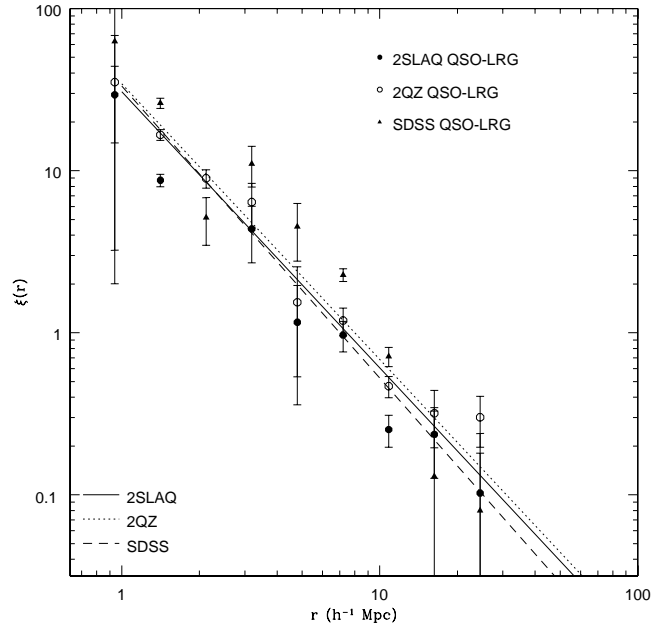}}
\caption{The real-space cross-correlation function, $\xi(r)$, results
 for our different samples. The dashed lines show the fits from the QSO-photometric LRG $w(\theta)$ measurements.}
\label{fig:xir}
\end{center}
\end{figure}

The QSO-2SLAQ (spectroscopic) LRG real-space results are shown in Fig. \ref{fig:xir}. In the same Figure we have also plotted the $\xi (r)$ fits from the QSO-photometric LRG $w(\theta )$ measurements, described in Section 4. All the samples seem to give consistent results although, as already mentioned, these $\xi (r)$ measurements from the spectroscopic samples are very noisy and no significant conclusions can be drawn. Finally, Table \ref{fig:table_xir} shows the $r_0$ and $\gamma$ values from the fits to the spectroscopic samples, on scales of 5$\leq r\leq 25$h$^{-1}$Mpc.

\begin{figure}
\begin{center}
\centerline{\epsfxsize = 9.0cm
\epsfbox{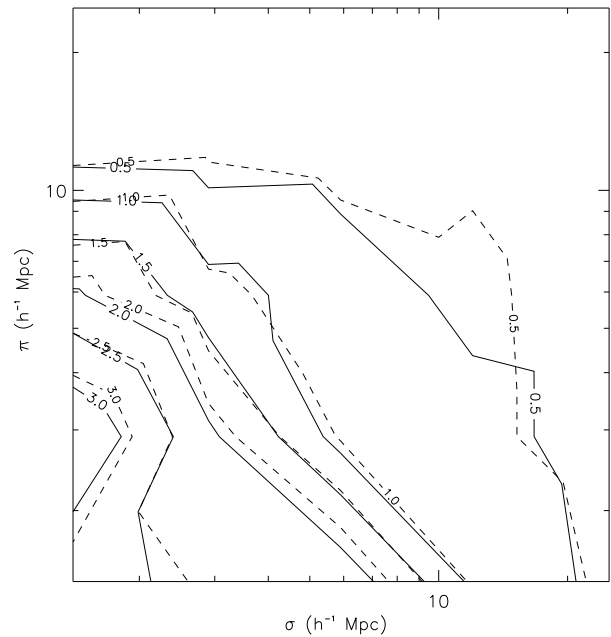}}
\caption{A comparison between our 2SLAQ QSO-2SLAQ LRG $\xi(\sigma, \pi)$ (solid line) results and the results using our Model I (dashed line). As we can see, model I is in very good agreement with the data both on small and large scales.}
\label{fig:xisigpi1}
\end{center}
\end{figure}

\section{Contraints on $\beta$ from redshift-space distortions}

\subsection{The $\xi(\sigma, \pi)$  cross-correlation function}

As already noted, QSO-LRG velocity dispersion and their dynamical infall into higher density regions are two mechanisms that distort the spherically symmetric clustering pattern in real space. We measure the $\xi(\sigma, \pi)$ cross-correlation function in the case of 2SLAQ, 2QZ and SDSS QSOs with the (spectroscopic) 2SLAQ LRGs. The purpose is to use the shape of the $\xi (\sigma, \pi)$ and our previous $\xi$ measurements in order to model the redshift-space distortions and put constraints on $\beta$. Fig. \ref{fig:xisigpi1} shows a comparison between our 2SLAQ QSO-2SLAQ LRG $\xi(\sigma, \pi)$ (solid line) results and the results using our model I (dashed line, see below). Model I is in very good agreement with the data both on small and large scales. 

\subsection{Description of $\xi(\sigma, \pi)$ models}

In order to place constraints on the infall parameter we model the redshift-space distortions measured above. First, we define our models for bias, $b$, and infall parameter $\beta$,

\begin{equation}
\xi _{mm}=\frac{\xi _{QL}}{b_Qb_L} 
\end{equation}
where $b_Q$ and $b_L$ are the QSO and LRG bias, respectively, given by

\begin{equation}
b_Q\simeq \frac{\Omega _m^{0.6}}{\beta _Q},  b_L\simeq \frac{\Omega _m^{0.6}}{\beta _L}
\label{eqn:bias}
\end{equation}
where $\beta _Q$ and $\beta _L$ are the QSO and LRG infall parameter, respectively and $\Omega _m$ is given in a flat universe as 

\begin{equation}
\Omega _m(z)=\frac{\Omega _m^0(1+z)^3}{\Omega _m^0(1+z)^3+\Omega
 _{\Lambda}^0}.
\label{eqn:omegamz}
\end{equation}

In our analysis in this Section we shall use two models. \textbf{Model I} is the one used in da $\hat{A}$ngela et al. (2008) modified accordingly to match our cross-correlation analysis, instead of the auto-correlation (QSO-QSO) which was used in that study. 

In general, the power spectrum in real and redshift space is given by (Kaiser 1987)

\begin{equation}
P_s(k)=(1+\beta (z) \mu _k ^2)^2P_r(k)
\label{eqn:powerspec}
\end{equation}
and the similar relation between $\xi (r)$ and $\xi (s)$ is (Hamilton 1993) 

\begin{equation}
\xi (r)=\frac{\xi(s)}{1+\frac{2}{3}\beta (z)+\frac{1}{5}\beta (z)^2}
\label{eqn:xir_xis}
\end{equation}
where $P_s(k)$ is the power-spectrum in redshift-space, $P_r(k)$ is the power-spectrum in real-space, and $\mu _k$ is the cosine of the angle between the wavevector $\textbf {k}$ and the line-of-sight. Equation \ref{eqn:powerspec} can also take the form,

\begin{eqnarray}
\xi (\sigma ,\pi ) & = & \left(1+\frac{2}{3}\beta (z)+\frac{1}{5}\beta
 (z)^2\right)\xi _0(r)P_0(\mu ) \nonumber \\
& &  \left(-\frac{4}{3}\beta (z)+\frac{4}{7}\beta(z)^2\right)\xi _2(r)P_2(\mu )
 \nonumber \\
& &  +\frac{8}{35}\beta (z)^2 \xi _4(r)P_4(\mu)
\label{eqn:model2}
\end{eqnarray}
$\mu$ is now the cosine of the angle between $r$ and $\pi$, $P_l(\mu)$ are the Legendre polynomials of order $l$ and $\xi _0(r)$, $\xi _2(r)$ and $\xi _4(r)$ being the monopole, quadrupole and hexadecapole components of the linear $\xi (r)$. 

In our analysis there are two infall parameters, $\beta _Q$ and $\beta _L$, one for QSOs and one for LRGs. Therefore, equations \ref{eqn:xir_xis}, \ref{eqn:powerspec} and \ref{eqn:model2} (Y. P. Jing, priv. communication) should be modified as follows: 

\begin{equation}
\xi (r)=\frac{\xi(s)}{1+\frac{1}{3}(\beta _Q (z)+\beta _L
 (z))+\frac{1}{5}\beta _Q (z)\beta _L (z)}
\label{eqn:kaiser_boost}
\end{equation}

\begin{equation}
P_s(k)=(1+\beta _Q (z) \mu _k ^2)(1+\beta _L (z) \mu _k ^2)P_r(k)
\end{equation}

\begin{eqnarray}
\xi (\sigma ,\pi ) & = & \left(1+\frac{1}{3} [\beta _Q (z)+\beta _L(z)]
 +\frac{1}{5}\beta _Q (z)\beta _L (z)\right)\xi _0(r)P_0(\mu ) \nonumber \\
& &  \left(-\frac{2}{3}[\beta _Q (z)+\beta _L(z)]+\frac{4}{7}\beta _Q(z)\beta
 _L(z)\right)\xi _2(r)P_2(\mu ) \nonumber \\
& &  +\frac{8}{35}\beta _Q(z)\beta _L(z) \xi _4(r)P_4(\mu)
\label{eqn:model_mod}
\end{eqnarray}
The problem with this formalism is that the model only constrains the sum of the infall parameters, i.e. $\beta _Q + \beta _L$, and not to each of them individually. So, in what follows we keep $\beta _L$ constant using the value found by Ross et al. (2007), $\beta _L=0.45\pm0.05$ and constrain the QSO infall parameter $\beta _Q$.

The magnitude of the elongation along the $\pi$-direction of the $\xi(\sigma, \pi )$ plot caused by the peculiar velocity of the object is denoted by $\langle \omega _z^2\rangle ^{1/2}$, which can be expressed by a Gaussian (Ratcliffe et al. 1996), as

\begin{equation}
f(\omega
 _z)=\frac{1}{\sqrt{2\pi}\langle \omega_z^2\rangle ^{1/2}}exp(-\frac{1}{2}\frac{|\omega_z|^2}{\langle \omega_z^2\rangle ^{1/2}})
\label{eqn:f_vel_dis}
\end{equation}
To include the small scale redshift-space effects due to the random motions of galaxies, we convolve the $\xi (\sigma, \pi)$ model with the peculiar velocity distribution, given by equation \ref{eqn:f_vel_dis}. Then, $\xi (\sigma, \pi)$ is given by

\begin{equation}
\xi (\sigma, \pi)=\int_{-\infty}^{+\infty}{\xi ^\prime(\sigma ,\pi
 -\omega _z(1+z)/H(z))f(\omega _z)\rm{d}\omega _z}
\end{equation}
where $\xi ^\prime(\sigma ,\pi -\omega _z(1+z)/H(z))$ is given by equation \ref{eqn:model2}.

Finally, we exploit the Alcock$-$Paczynski effect ( Alcock \& Paczynski 1979) which says that the ratio of observed angular size to radial size varies with cosmology (isotropic clustering). If we assume that the cluster is isotropic, then we can constrain the cosmological parameters by requiring that they produce equal tangential and radial sizes. In our case the angular and the radial size are $\sigma$ and $\pi$. Keeping the meanings of test and assumed cosmology as used by da $\hat{A}$ngela et al. (2008) and following their fitting procedure we obtain our results (see below). In particular, we use the $\gamma$ values from our fits to $\xi (s)$, let $r_0$ and the velocity dispersion vary as free parameters and compute the $\chi ^2$ values for each $\Omega _m^0-\beta (z)$ pair.

The second model (\textbf{model II}) is as follows. $\xi (\sigma , \pi )$ is now defined as (Peebles 1980, Hoyle 2000) 

\begin{equation}
1+\xi (\sigma, \pi)=\int_{-\infty}^{+\infty}{(1+\xi (r))f(\omega
 _z)\rm{d}\omega _z}
\label{eqn:sigma2}
\end{equation}
where the $f(\omega _z)$ is given, as before, by equation \ref{eqn:f_vel_dis}. Next we introduce the infall velocity of the galaxies, $\upsilon (r_z)$, as a function of the real-space separation along the $\pi$ direction, $r_z$. This can be derived from the equation of conservation of particle pairs, which are within a comoving separation $r$ from a mass particle, i.e. (Peebles 1980)

\begin{equation}
\frac{\delta}{\delta t}\int_{0}^{r}{\chi ^2 \xi
 ^m(x,t)\rm{dx}+\frac{1}{a(t)}r^2(1+\xi ^m(r,t))\upsilon(r,t)}=0
\end{equation}
where $a(t)$ is the scale factor. Assuming that $\xi (r)$ is described by a power law model, we solve the above equation to find the infall velocity of the particles  

\begin{equation}
\upsilon (r_z)=-\frac{2}{3-\gamma}\Omega
 _m(z)^{0.6}H(z)r_z\frac{\xi_ {QL}(r)}{b_Qb_L+\xi _{QL}(r)}
\end{equation}
We now modify equation \ref{eqn:sigma2} to include the effects of the
 bulk motions

\begin{equation}
1+\xi (\sigma, \pi)=\int_{-\infty}^{+\infty}{(1+\xi (r))f(\omega
 _z(1+z)-\upsilon (r_z))\rm{d}\omega _z}
\label{eqn:sigma3}
\end{equation}

This is the model II $\xi (\sigma ,\pi )$; we then follow the same implementation of the ``Alcock-Paczynski'' effect and fitting procedure as for model I. As in model I, we keep $b_L$ constant, using the same value as above (Ross et al. 2007), i.e. $\beta _L=0.45\pm0.05$ and $b_L=1.66\pm0.35$.

\subsection{Results}

\begin{figure}
\centerline{\epsfxsize = 5.0cm
\epsfbox{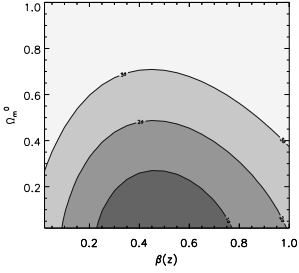}}
\caption{Likelihood contours of $\Omega _m^0-\beta (z=0.55)$ for 2SLAQ
 QSO$-$2SLAQ LRGs, using model I. A $\Lambda$CDM cosmology is assumed
 along with a model where $r_0=8.2$, $\gamma=1.6$ with
 $\langle w_z^2\rangle ^{1/2}=630$kms$^{-1}$. The best fit value is $\beta _Q
 (z=0.55)=0.45_{-0.22}^{+0.32}$.}
\label{fig:2slaq_1}

\end{figure}

\begin{figure}
\centerline{\epsfxsize = 5.0cm
\epsfbox{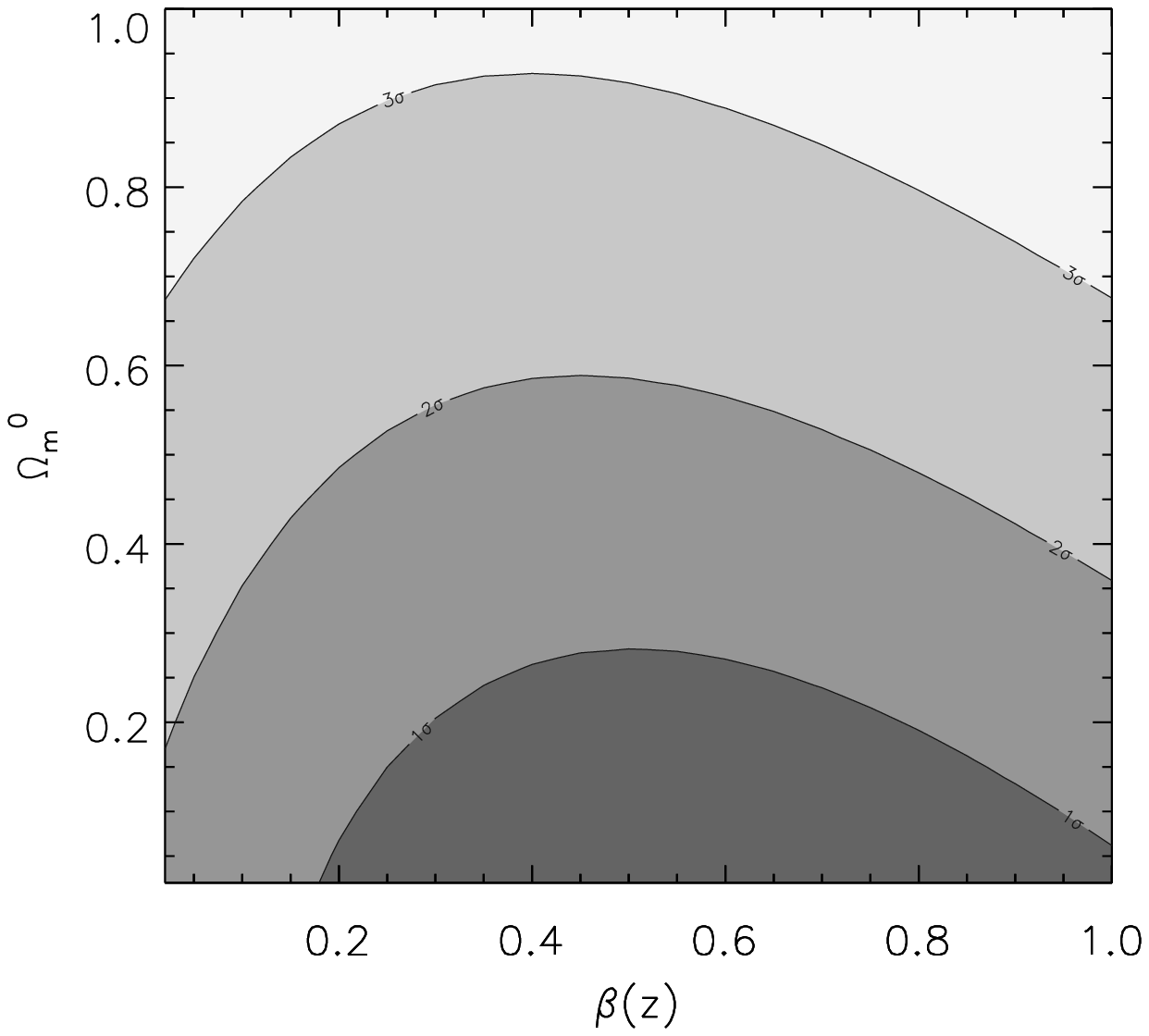}}
\caption{Likelihood contours of $\Omega _m^0-\beta (z=0.55)$ for 2QZ
 QSO$-$2SLAQ LRGs, using model I. A $\Lambda$CDM cosmology is assumed
 along with a model where $r_0=8.0$, $\gamma=1.7$ with
 $\langle w_z^2\rangle ^{1/2}=560$kms$^{-1}$. The best fit value is $\beta _Q(z=0.55)=0.55_{-0.38}^{+0.53}$.}
\label{fig:2qz_1}
\end{figure}

\begin{figure}
\centerline{\epsfxsize = 5.0cm
\epsfbox{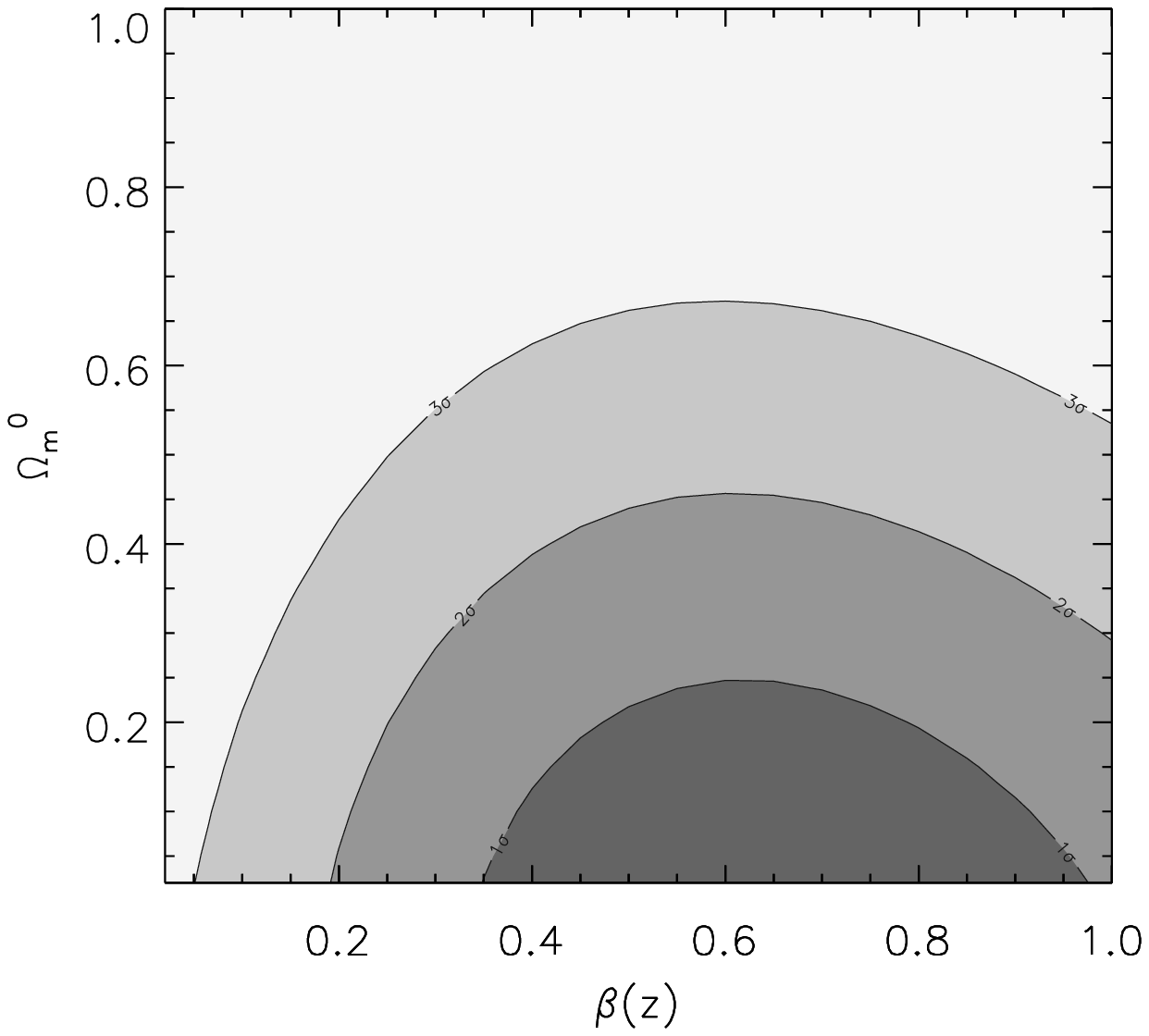}}
\caption{Likelihood contours of $\Omega _m^0-\beta (z=0.55)$ for SDSS
 QSO$-$2SLAQ LRGs, using model I. A $\Lambda$CDM cosmology is assumed
 along with a model where $r_0=7.5$, $\gamma=1.8$ with
 $\langle w_z^2\rangle ^{1/2}=670$kms$^{-1}$. The best fit value is $\beta _Q
 (z=0.55)=0.60_{-0.25}^{+0.35}$.}
\label{fig:sdss_1}
\end{figure}

\begin{figure}
\centerline{\epsfxsize = 5.0cm
\epsfbox{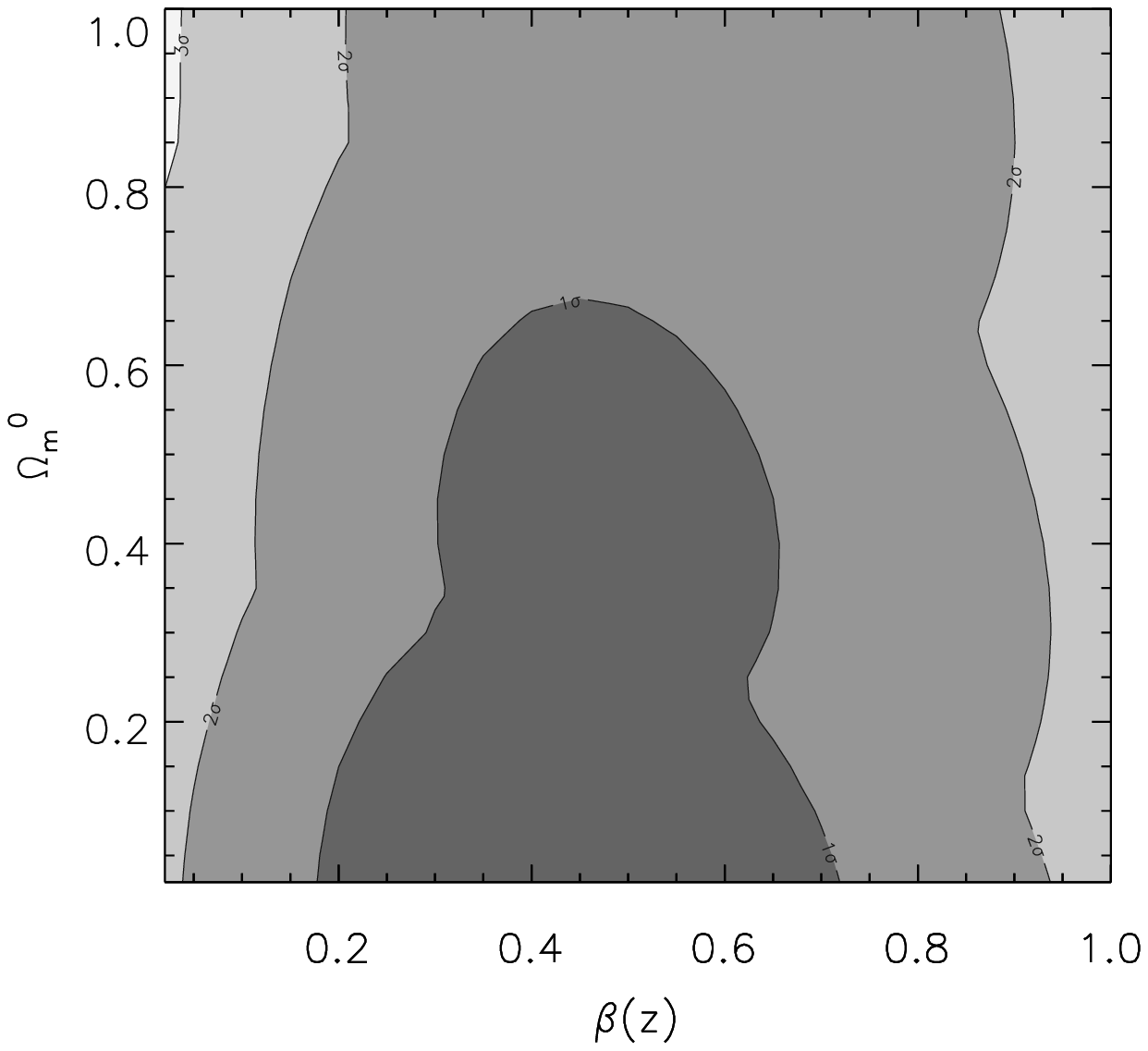}}
\caption{Likelihood contours of $\Omega _m^0-\beta (z=0.55)$ for 2SLAQ
 QSO$-$2SLAQ LRGs, using model II. A $\Lambda$CDM cosmology is assumed
 along with a model where $r_0=8.2$, $\gamma=1.6$ with
 $\langle w_z^2\rangle ^{1/2}=750$kms$^{-1}$. The best fit value is $\beta _Q
 (z=0.55)=0.40_{-0.22}^{+0.32}$.}
\label{fig:2slaq_2}
\end{figure}

\begin{figure}
\centerline{\epsfxsize = 5.0cm
\epsfbox{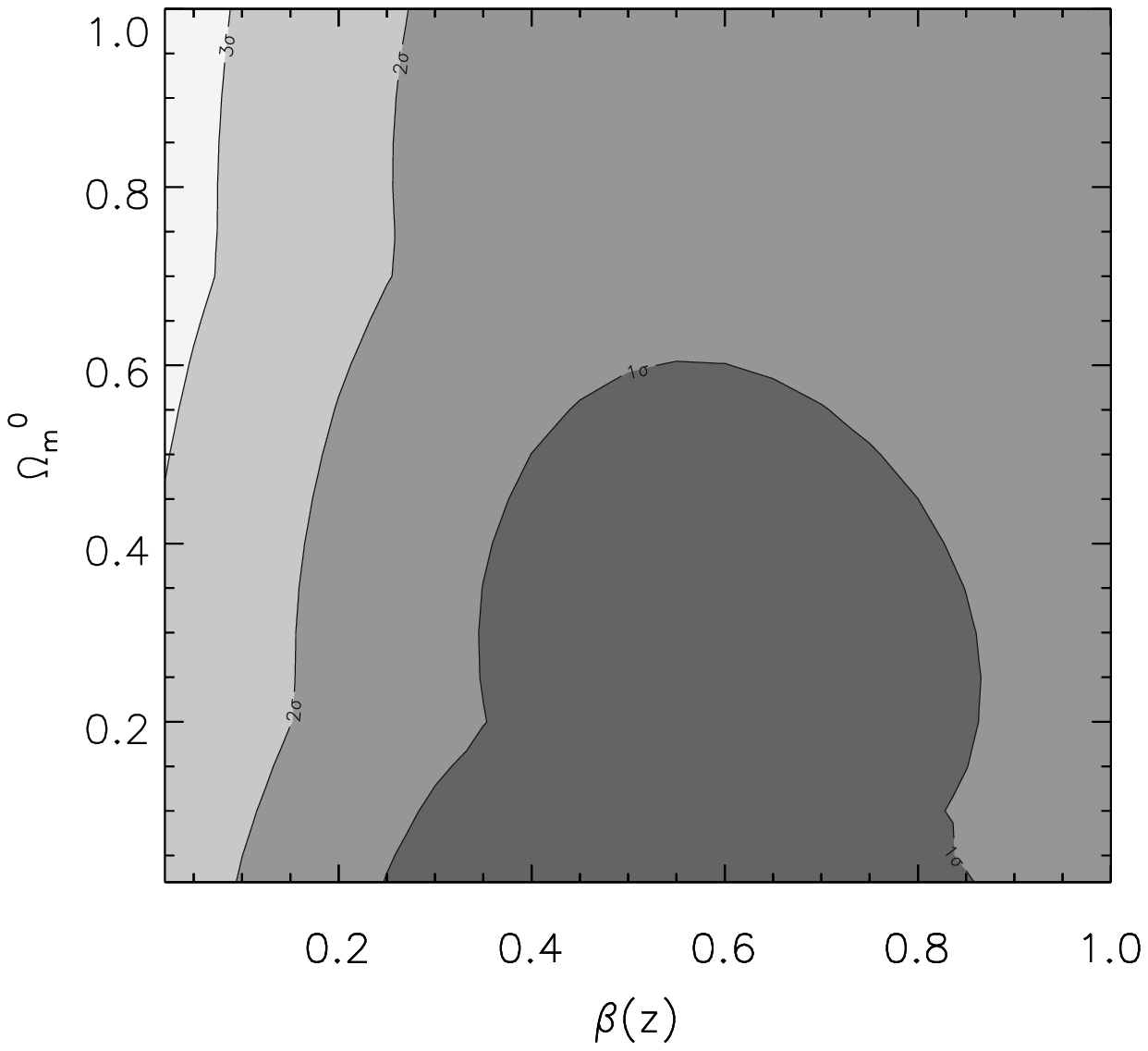}}
\caption{Likelihood contours of $\Omega _m^0-\beta (z=0.55)$ for 2QZ
 QSO$-$2SLAQ LRGs, using model II. A $\Lambda$CDM cosmology is assumed
 along with a model where $r_0=8.0$, $\gamma=1.7$ with
 $\langle w_z^2\rangle ^{1/2}=710$kms$^{-1}$. The best fit value is $\beta _Q
 (z=0.55)=0.60_{-0.35}^{+0.25}$.}
\label{fig:2qz_2}
\end{figure}

\begin{figure}
\centerline{\epsfxsize = 5.0cm
\epsfbox{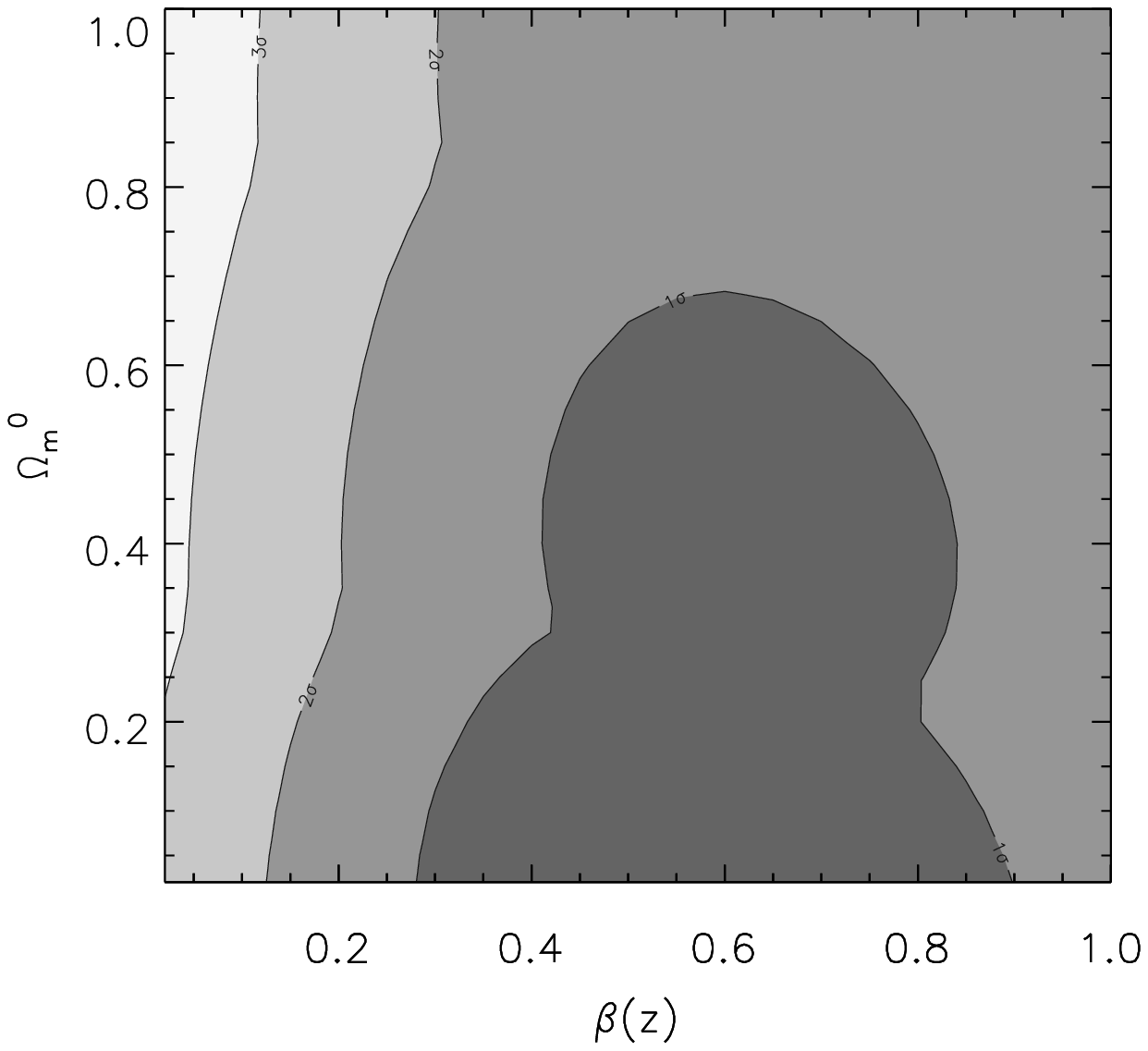}}
\caption{Likelihood contours of $\Omega _m^0-\beta (z=0.55)$ for SDSS
 QSO$-$2SLAQ LRGs, using model II. A $\Lambda$CDM cosmology is assumed
 along with a model where $r_0=7.5$, $\gamma=1.8$ with
 $\langle w_z^2\rangle ^{1/2}=710$kms$^{-1}$. The best fit value is $\beta
 (z=0.55)=0.65_{-0.37}^{+0.25}$.}
\label{fig:sdss_2}
\end{figure}

We now use our $\xi (\sigma, \pi)$ measurements from the previous Section and the $s_0$ and $\gamma$ values from the $\xi (s)$ fits shown in Table \ref{table:spectroscopic} to put constraints on $\beta_Q$ and $b_Q$ and $\langle \omega_z^2\rangle ^{1/2}$ for each of the QSO samples. The results are shown in Table \ref{Table:beta_b_redz} and in Figures \ref{fig:2slaq_1}-\ref{fig:sdss_2}.

Comparing the results from both models from the three different QSO samples associated with the same spectroscopic 2SLAQ LRG sample, we note that model I gives slightly lower QSO-LRG velocity dispersions than model II, but both are consistent ($\sim$620kms$^{-1}$ and $\sim727$kms$^{-1}$) with the expected $\langle \omega_z^2\rangle ^{1/2}\simeq$728kms$^{-1}$, produced by adding in quadrupole QSO and LRG velocity dispersions from previous QSO-QSO and LRG-LRG studies, which gave 800kms$^{-1}$ and 330kms$^{-1}$, respectively. Comparison of $b_Q$ between the different samples (for both models) shows that the results are in good statistical agreement (Table \ref{Table:beta_b_redz}). The best way to summarise the results is to average them. All six measurements (the three samples and the two models) give an average of $\beta _Q=0.55\pm0.10$ and $b_Q=1.4\pm0.2$ at z=0.55, which is consistent with the values found by da $\hat{A}$ngela et al. (2008), $\beta _Q=0.60_{-0.11}^{+0.14}$ and $b_Q=1.5\pm0.2$ at z=1.4 and $b_Q=1.1\pm0.2$ at $z\simeq0.6$ (see their Fig. 13).

\section{QSO bias and halo masses}  

\subsection{QSO-LRG clustering dependence on luminosity}

Previous attempts to study the dependence of clustering on luminosity, were not successful because of the redshift-luminosity degeneracy, as higher luminosity QSOs reside at higher redshifts. Here, combining QSOs with large LRG samples we get the statistical power to break this degeneracy. In this Section, we shall try to examine if and how the QSO-LRG clustering depends on QSO luminosity, at fixed redshift. 

We first estimate the average absolute magnitude of each of our QSO samples, as follows:

\begin{equation}
M_{b_J}=b_J-25-5\rm{log}_{10}\it d_L+2.5(1+\alpha ^\prime)\rm{log}_{10}(1+z)
\end{equation}
where $M_{b_J}$ is the absolute magnitude of each QSO, $b_J$ (or $g$) is its apparent magnitude , $d_L$ is the luminosity distance (Mpc) that corresponds to the redshift $z$ and the last term is the k-correction, where we have assumed a QSO spectral index $\alpha^\prime=-1.0$. We should note here that we treat the $b_J$ band (2QZ) as being equivalent to the $g$ band (2SLAQ, SDSS; Richards et al. 2005). 

To check if there is a dependence of QSO-LRG clustering on luminosity we use the integrated correlation function, as a more robust statistical tool (see da $\hat{A}$ngela et al. 2008). We calculate it up to scales of 20h$^{-1}$Mpc and normalise the result to the volume contained in a sphere with radius of 20h$^{-1}$Mpc:

\begin{equation}
\xi _{20}=\frac{3}{20^3}\int_0^{20}{\xi (s)s^2ds}
\label{eqn:integ_xis}
\end{equation}
The choice of the radius has been made for two reasons. The 20h$^{-1}$Mpc scale is large enough to apply linear theory (Croom et al. 2005), and redshift-space distortions (finger of god or redshift errors) do not significantly affect the measurements. 

\begin{figure*}
\begin{center}
\centerline{\epsfxsize = 9.0cm
\epsfbox{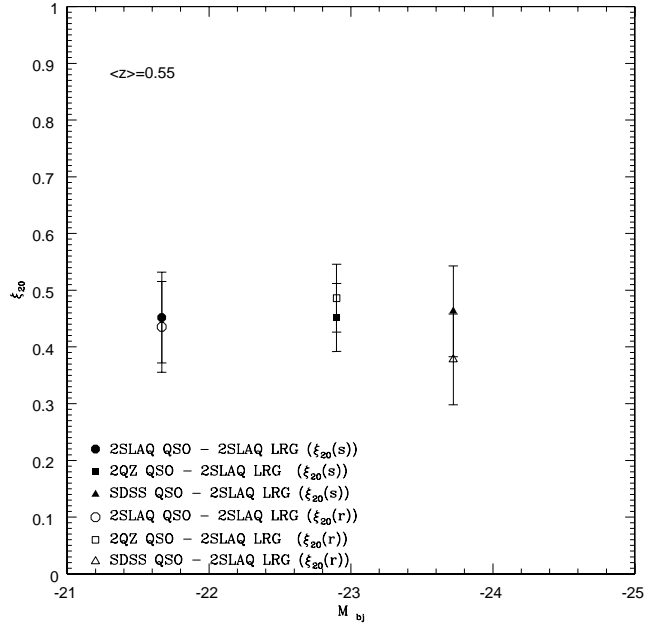}}
\caption{$\xi _{20}$ cross-correlation measurements of the three QSO samples with 2SLAQ LRGs. Filled symbols show the results using spectroscopic samples and open symbols using photometric (LRG) samples. }
\label{fig:xi_bar_2slaq}
\end{center}
\end{figure*}

\begin{figure*}
\begin{center}
\centerline{\epsfxsize = 9.0cm
\epsfbox{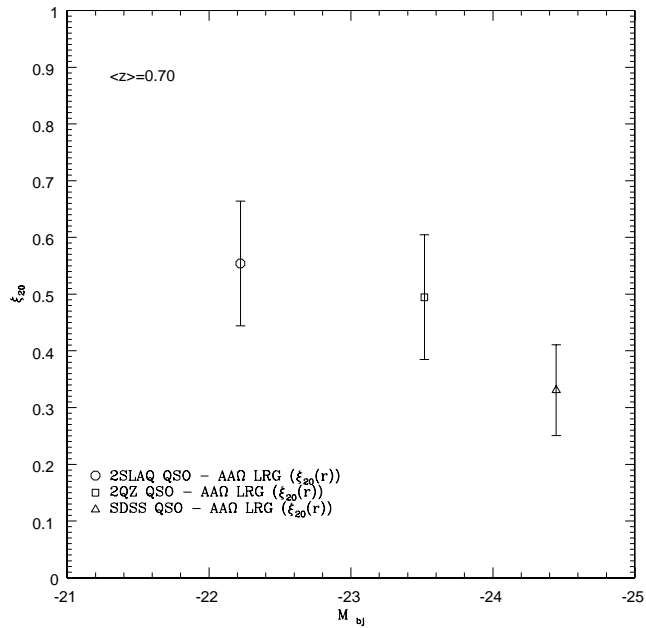}}
\caption{$\xi _{20}$ cross-correlation measurements from the three QSO samples with (photometric) AAOmega LRGs.}
\label{fig:xi_bar_aao}
\end{center}
\end{figure*}

In the case of the QSOs-2SLAQ (spectroscopic) LRGs we have estimated $\xi _{20}$ via equation \ref{eqn:integ_xis} using the $\xi (s)$ measurements shown in Fig. \ref{fig:xis_all1}. For the QSO-2SLAQ (photometric) and AAOmega LRGs we have substituted the $\xi (s)$ in equation \ref{eqn:integ_xis} with the $\xi (r)$ fits shown in Fig. \ref{fig:Limber_all}. The results using 2SLAQ LRGs (spectroscopic and photometric) are shown in Fig. \ref{fig:xi_bar_2slaq} and using AAOmega LRGs are shown in Fig. \ref{fig:xi_bar_aao}. No conclusion can be drawn about the redshift evolution of the QSO-LRG clustering as the average redshift of the samples is too restricted. Although the $\xi _{20}$ results using the AAOmega LRGs, show some indications that bright QSOs (SDSS) cluster less with LRGs than faint QSOs (2SLAQ), the results using the 2SLAQ LRG samples (both spectroscopic and photometric) stay statistically constant, thus confirming the results in the previous Sections, that the QSO-LRG clustering is independent of QSO luminosity.

\subsection{QSO bias}

\begin{figure*}
\begin{center}
\centerline{\epsfxsize = 9.0cm
\epsfbox{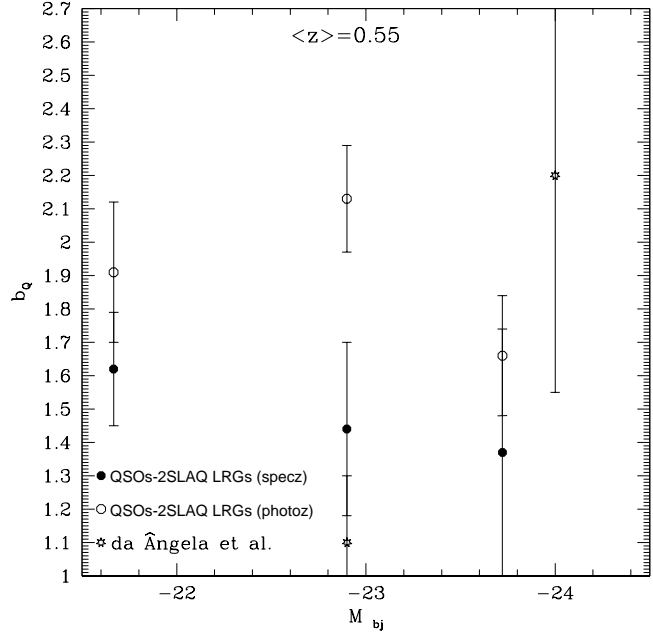}}
\caption{Measurement of the QSO bias, $b_Q$, for QSOs and 2SLAQ LRG samples. For consistency, spectroscopic samples use $\xi _{20}$ from $w_p(\sigma)$ in Table \ref{table:biases_qso} rather than the $\xi (s)$ values shown in Fig. \ref{fig:xi_bar_2slaq}. Stars show the two points taken from Fig. 13 of da $\hat{A}$ngela et al. (2008). The fainter one is at $\langle z\rangle \simeq 0.6$ and the brighter at $\langle z\rangle \simeq 0.7$.}
\label{fig:biases_qso_1}
\end{center}
\end{figure*}

\begin{figure*}
\begin{center}
\centerline{\epsfxsize = 9.0cm
\epsfbox{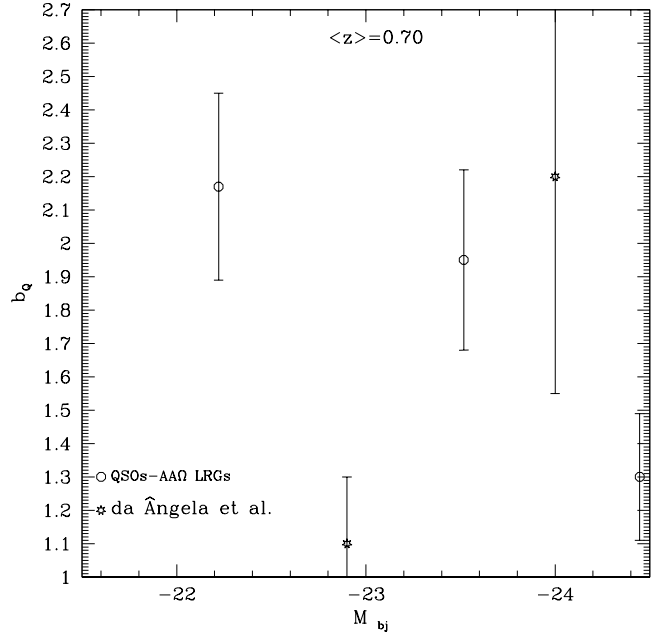}}
\caption{Measurement of the QSO bias, $b_Q$, for QSOs and the (photometric) AAOmega LRG sample. Stars show the two points taken from Fig. 13 of da $\hat{A}$ngela et al. (2008). The fainter one is at $\langle z\rangle \simeq 0.6$ and the brighter at $\langle z\rangle \simeq 0.7$.}
\label{fig:biases_qso_2}
\end{center}
\end{figure*}

Having tested the luminosity dependence of the QSO-LRG clustering, we now investigate the dependence of the QSO bias on the luminosity. Assuming that this bias is independent of scale, we calculate the QSO bias using:

\begin{equation}
b_Qb_L=\frac{\xi _{QL}(r)}{\xi _{mm}}\Rightarrow b_Q\approx\frac{1}{b_L}\frac{\xi _{QL}(r,20)}{\xi_{mm}(r,20)}
\label{eqn:bias_rho}
\end{equation}
where $\xi _{\rho}$ is the matter real-space correlation function, averaged in 20h$^{-1}$Mpc spheres. In the case of QSO-photometric LRGs we use our $\xi (r,20)$ measurements as estimated before. For consistency, for the QSO-2SLAQ (spectroscopic) LRGs, we use our $w_p(\sigma)$ measurements (shown in Table \ref{table:biases_qso}) since they are less noisy than those for $\xi (r)$. To estimate $\xi _{mm}$ we use the values as estimated by da $\hat{A}$ngela et al. 2008. So in the case of QSO-AAOmega LRGs $\xi _{mm}=0.11$ at ($z\approx0.7$), and in the case of QSO-2SLAQ LRGs $\xi _{mm}=0.12$ ($z\approx0.55$). Finally, we calculate the LRG bias for the AAOmega and 2SLAQ, based on the results shown in Table 4 of Ross et al. (2007) (AAOmega: $r_0=9.03\pm0.93$ and $\gamma=1.73\pm0.08$, 2SLAQ: $r_0=7.45\pm0.35$ and $\gamma=1.72\pm0.06$). We find, $b_{L(AA\Omega)}=2.35\pm0.20$ and $b_{L(2SLAQ)}=1.90\pm0.08$. The latter is in reasonable agreement with the value found by Ross et al. (2007), from redshift-space distortions, $b_{L(2SLAQ)}=1.66\pm0.35$. The derived QSO bias values for each case are shown in Table \ref{table:biases_qso} (as well as the corresponding $\beta _Q$ values) and in Figures \ref{fig:biases_qso_1} and \ref{fig:biases_qso_2}. 

Comparing the values for the QSO biases from the different samples, we note that the QSO biases using 2SLAQ LRG samples show indications for luminosity dependent QSO bias, in the sense that $b_Q$ reduces for higher luminosity samples, at least in the case of spectroscopic 2SLAQ LRGs. The same pattern is repeated when using AAOmega LRG samples. The spectroscopic samples yield lower $b_Q$ values than the photometric samples. This is due to the fact that the amplitude of the $\xi (r)$ measurements of the photometric samples is higher than the amplitude of the $w_p(\sigma)$ measurements of the spectroscopic samples (see Fig. \ref{fig:projected1}). Combining the 2SLAQ (photometric and spectroscopic) samples with the photometric AAOmega samples we find $b_Q=1.90\pm0.16$, $b_Q=1.85\pm0.23$ and $b_Q=1.45\pm0.11$, for 2SLAQ, 2QZ and SDSS QSOs, respectively. Comparing now the values for the QSO bias from the spectroscopic 2SLAQ LRG samples with those obtained in Section 7.3, we note that the amplitude results, give an average of $b_Q=1.5\pm0.1$ which is in very good agreement with the average of $b_Q=1.4\pm0.2$ obtained from the redshift-space distortion results. Our measurements give an overall QSO bias of $b_Q\approx1.5$ at $z=0.55$ and $M_{b_J}\approx-23$.

In Figures \ref{fig:biases_qso_1} and \ref{fig:biases_qso_2} we have also plotted two points (stars), that are at low redshifts, taken from Fig. 13 of da $\hat{A}$ngela et al. (2008). The one with $M_{b_J}\simeq-24.0$ at $z\simeq0.7$ is in statistical  agreement with our $b_Q$ values from the AAOmega LRG samples, at the same mean redshift and brightness. The second one with $M_{b_J}\simeq-22.9$ at $z\simeq0.6$ is lower than our $b_Q$ values from 2SLAQ LRG samples, at $z=0.55$, but statistically not rejected by them (at least not by those from the spectroscopic samples). The overall impression is that our $b_Q\approx1.5$ at $z=0.55$ is in agreement with the values found by da $\hat{A}$ngela et al. (2008), $b_Q=1.5\pm0.2$ at $z=1.4$ and slightly higher than $b_Q=1.1\pm0.2$ found at $z\simeq0.6$.

\subsection{Dark Matter Halo Mass}

Since the bias of Dark Matter Halos is correlated to their mass (Mo \& White 1996), we shall attempt to measure this mass ($M_{DMH}$). In our analysis we shall follow da $\hat{A}$ngela et al. and Croom et al. and assume an ellipsoidal collapse model, described by Sheth et al. (2001). The bias and the $M_{DMH}$ are related via

\begin{eqnarray}
b(M_{DMH})& = & 1+\frac{1}{\alpha ^{0.5}\delta _c(z)}[\alpha
 ^{0.5}(\alpha \nu ^2)+\alpha ^{0.5}b(\alpha \nu ^2)^{1-c} \nonumber \\
& & -\frac{(\alpha \nu ^2)^c}{\alpha \nu ^2+b(1-c)(1-\frac{c}{2})}]
\end{eqnarray}
where $\alpha=0.707$, $b=0.5$ and $c=0.6$. $\nu$ is defined as $\nu=\delta _c(z)/\sigma(M_{DMH},z)$, with $\delta _c$ to be the critical density for collapse, given by, $\delta _c=0.15(12\pi)^{\frac{2}{3}}\Omega_m(z)^{0.0055}$ (Navarro et al. 1997). $\sigma(M_{DMH},z)=\sigma(M_{DMH})G(z)$, where $\sigma(M_{DMH})$ is the rms fluctuation of the density field on the mass scale with value $M_{DMH}$ and $G(z)$ is the linear growth factor (Peebles 1984). The $\sigma (M_{DMH})$ can then calculated as

\begin{equation}
\sigma(M_{DMH})^2=\frac{1}{2\pi ^2}\int_0^\infty{k^2P(k)w(kr)^2dk}
\end{equation}
with P(k) to be the power spectrum of density perturbations and $w(kr)$
 is the Fourier transform of a spherical top hat, which is given by
 (Peebles 1980):

\begin{equation}
w(kr)=3\frac{\rm{sin}\it(kr)-kr\rm{cos}\it(kr)}{(kr)^3}
\end{equation}
where the radius and mass are related through

\begin{equation}
r=\left(\frac{3M_{DMH}}{4\pi \rho _0}\right)^\frac{1}{3}
\end{equation}
where $\rho _0$ is the present mean density of the Universe, given by
 $\rho _0=\Omega_m^0\rho_{crit}^0=2.78\times10^{11}\Omega _m^0h^2M_{\sun}
 Mpc^{-3}$. The power spectrum used in our analysis has the linear form,
 $P(k)=P_0T(k)^2k^n$, with $P_0$ to be a normalisation parameter which
 depends on $\sigma_8$ and T(k) is the transfer function (Bardeen et al.
 1986).

\begin{figure*}
\begin{center}
\centerline{\epsfxsize = 9.0cm
\epsfbox{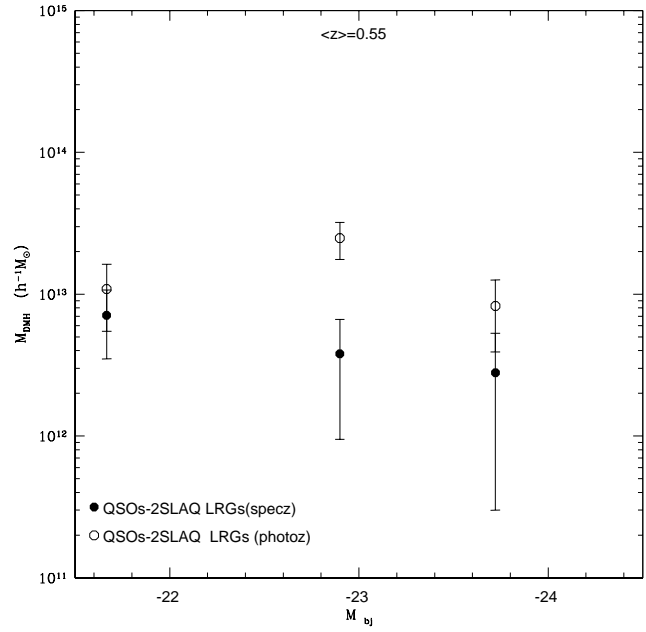}}
\caption{Measurement of the $M_{DMH}$, for different QSO and 2SLAQ LRG samples.}
\label{fig:halo_qso_1}
\end{center}
\end{figure*}

\begin{figure*}
\begin{center}
\centerline{\epsfxsize = 9.0cm
\epsfbox{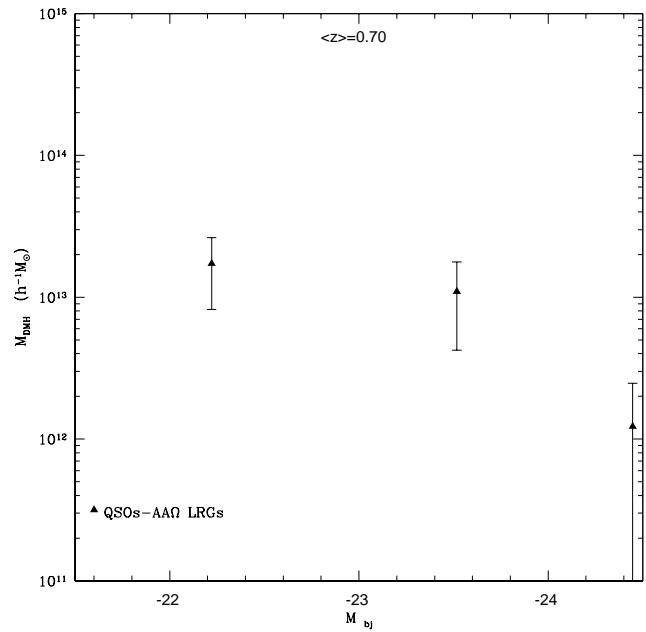}}
\caption{Measurement of the $M_{DMH}$, for different QSO and AAOmega LRG samples.}
\label{fig:halo_qso_2}
\end{center}
\end{figure*}

The results are shown in Figures \ref{fig:halo_qso_1} and \ref{fig:halo_qso_2}. Once again, although for the AAOmega LRG samples the derived QSO halo masses show indications of increasing as we move to fainter QSO samples, in the case of 2SLAQ (photometric and spectroscopic) LRG samples halo masses stay statistically constant. The average value is $M_{DMH}=10^{13}h^{-1}M_{\sun}$. Comparing now this result with those from other authors (Croom et al. 2005, da $\hat{A}$ngela et al. 2008), we note that their $M_{DMH}$ estimates are generally lower than ours ($\sim3\times10^{12}h^{-1}M_{\sun}$) although at higher redshifts ($z=1.4$). They also find that the hosts of QSOs have the same mass at all redshifts, thus rejecting cosmologically long-lived QSO models. Our higher masses at $z=0.55$ may be more consistent with the long-lived predictions of $6\times10^{14}h^{-1}M_{\sun}$ at $z=0$ and $10^{13}h^{-1}M_{\sun}$ at $z\simeq0.5$. The caveat is that for our measurements we need to use a value for $b_L$ in order to derive $b_Q$.

\section{Discussion $+$ Conclusions}

In this paper we have performed an analysis of the clustering of QSOs with LRGs. For this purpose, we first used the 2-point angular cross-correlation function, $w(\theta)$, and measured the cross-correlation between 2SLAQ and AAOmega LRGs and different luminosity QSOs. The results show that there is little cross-correlation dependence on QSO luminosity.

Next, we measured the redshift-space cross-correlation function. We again see no QSO-LRG clustering dependence on QSO luminosity, as all the QSO-spectroscopic LRG samples gave similar results. We used Limber's formula to fit $r_0$ to 2-D results. The fits for $r_0$ from 3-D $\xi (s)$ are in very good agreement with the fits to the 2-D $w(\theta)$ results. Then, we compared our QSO-LRG clustering with 2SLAQ LRG-LRG (Ross et al. 2007) and 2QZ+2SLAQ QSO-LRG (da $\hat{A}$ngela et al. 2008) clustering results. On small scales, the QSO-QSO and QSO-LRG results appear flatter than the LRG-LRG results. As confirmed later by the $w_p(\sigma )/\sigma$ measurements, this appears to be due to the larger QSO redshift errors (broad-lines). On larger scales ($\geq 5$h$^{-1}$Mpc) the QSO-QSO correlation amplitude appears lower than the QSO-LRG and LRG-LRG amplitudes, suggesting that $b_Q\lesssim b_L$. The fractional errors on $\xi _{QL}$ are $\sim 50\%$ smaller than those on $\xi _{QQ}$ in same redshift range.

The results from the semi-projected cross-correlation function, $w_p(\sigma )/\sigma$, are in agreement with our $\xi (s)$ observations, yielding consistent $r_0$ and $\gamma$ values. The comparison with the 2SLAQ LRG-LRG (Ross et al. 2007) and 2QZ+2SLAQ QSO-LRG (da $\hat{A}$ngela et al. 2008) confirms our results from redshift-space, i.e. that the small-scale amplitude difference in $\xi (s)$ is due to the larger QSO redshift errors and that QSO-QSO clustering amplitude is lower than QSO-LRG and LRG-LRG amplitude. The real-space cross-correlation function, $\xi (r)$, also seems to agree with the $\xi (s)$ and $w_p(\sigma )/\sigma$ results, although it is noisier.

Then, we measured the $\xi (\sigma, \pi)$ cross-correlation function for all our (spectroscopic) samples and used the results to model the redshift-space distortions. For that, we used two models which gave consistent results with each other and between the different samples. The redshift-space distortions yielded an average of $\beta _Q=0.55\pm0.10$, $b_Q=1.4\pm0.2$ which is slightly higher than $b_Q=1.1\pm0.2$ at $z\simeq0.6$, from da $\hat{A}$ngela et al. (2008). Note that this latter result does not come from analysing the amplitude of $\xi _{QQ}$; there being too few QSO pairs to make redshift-distortion auto-correlation analysis viable at $z\sim 0.6$.

After calculating the average absolute magnitude of each QSO sample we measured the integrated cross-correlation function for each one of our QSO-LRG samples. No evidence was found for QSO-LRG clustering dependence on QSO luminosity. Then, using LRG biases as estimated by previous studies, we calculated the QSO biases. There were indications of luminosity dependence, in the sense that $b_Q$ may reduce as we move to brighter QSO samples. Our analysis yielded a $b_Q=1.5\pm0.1$ (at $M_{b_J}\approx-23$) which is in very good agreement with our result from redshift-space distortions.
 
Finally, using the relation between bias and $M_{DMH}$ suggested by Sheth et al. 2001, we calculated the corresponding masses of the QSO hosts. QSO halo masses were estimated to be $\sim10^{13}h^{-1}M_{\sun}$ at $z\approx0.55$. Our $M_{DMH}$ estimations are higher than those from Croom et al. (2005) and da $\hat{A}$ngela et al. (2008) (at $z=1.4$) and may be explained by long-lived QSO population models. Since the bias values are independent of QSO luminosity at fixed redshift, the halo masses are also independent of luminosity and this represents the main result of this paper.

\section{Acknowledgments}

Funding for the SDSS and SDSS-II has been provided by the Alfred P.
Sloan Foundation, the Participating Institutions, the National Science
Foundation, the U.S. Department of Energy, the National Aeronautics and
Space Administration, the Japanese Monbukagakusho, the Max Planck
Society, and the Higher Education Funding Council for England. The SDSS
Web Site is http://www.sdss.org.

The SDSS is managed by the Astrophysical Research Consortium for the
Participating Institutions. The Participating Institutions are the
American Museum of Natural History, Astrophysical Institute Potsdam,
University of Basel, Cambridge University, Case Western Reserve
University, University of Chicago, Drexel University, Fermilab, the
Institute for Advanced Study, the Japan Participation Group, John
Hopkins University, the Joint Institute for Nuclear Astrophysics, the
Kavli Institute for PArticle Astrophysics and Cosmology, the Korean
Scientist Group, the Chinese Academy of Sciences (LAMOST), Los Alamos
NAtional Observatory, the Max-Planck-Institute for Astronomy (MPIA),
 the
Max-Planck-Institute for Astrophysics (MPA), New Mexico State
University, Ohio State Univesity, University of Pittburgh, University
 of
Portsmouth, Princeton University, the United States Naval Observatory,
and tge UNiversity of Washington.

The 2dF QSO Redshift Survey (2QZ) was compiled by the 2QZ survey team
from observations made with the 2-degree Field on the Anglo-Australian
Telescope.
\vspace{10 mm}

\noindent
{\bf References}

\vspace{6 mm}

\noindent
Ballinger, W. E., Peacock, J. A., Heavens, A. F., 1996, MNRAS, 282, 877

\vspace{3 mm}

\noindent
Bardeen J.M., Bond J.R. Kaiser N., Szalay A. S., 1986, ApJ, 304, 45

\vspace{3 mm}

\noindent
Cannon R., et al., 2006, MNRAS, 372, 425

\vspace{3 mm}

\noindent
Coil A.L., Hennawi J.F., Newman J.A., Cooper M.C., Davis M. 2007, ApJ, 654, 115

\vspace{3 mm}

\noindent
Croom S. M., et al., 2001, MNRAS, 325, 483

\vspace{3 mm}

\noindent
Croom S. M., Boyle B. J., Loaring N. S., Miller L., Outram P. J.,
Shanks T., Smith R. J., 2002, MNRAS, 335, 456

\vspace{3 mm}

\noindent
Croom S. M., et al., 2004, MNRAS, 349, 1397

\vspace{3 mm}

\noindent
Croom S. M., et al., 2005, MNRAS, 356, 415

\vspace{3 mm}

\noindent
da $\hat{A}$ngela J., 2006, Ph.D. Thesis, Durham University

\vspace{3 mm}

\noindent
da $\hat{A}$ngela, J. et al., 2008, MNRAS, 383, 565D

\vspace{3 mm}

\noindent
da $\hat{A}$ngela, J., Outram, P. J., Shanks, T., 2005, MNRAS, 361, 879

\vspace{3 mm}

\noindent
da $\hat{A}$ngela, J., Outram, P. J., Shanks, T., Boyle, B. J., Croom,
 S. M., Loaring, N. S.; Miller, L., Smith, R. J. 2005, MNRAS, 360, 1040

\vspace{3 mm}

\noindent
Eisenstein D. J., et al., 2001, AJ, 122, 2267

\vspace{3 mm}

\noindent
Eisenstein D. J., et al., 2005, ApJ, 633, 560

\vspace{3 mm}

\noindent
Georgantopoulos I., Shanks T., 1994, MNRAS, 271, 773

\vspace{3 mm}

\noindent 
Hamilton A. J. S., 1993, ApJ, 417, 19

\vspace{3 mm}

\noindent 
Hawkins et al., 2003, MNRAS, 346, 78

\vspace{3 mm}

\noindent
Hoyle, F., Outram, P. J., Shanks, T., Boyle, B. J., Croom, S. M.,
 Smith, R. J., 2002, MNRAS, 332, 311

\vspace{3 mm}

\noindent
Kaiser N., 1987, MNRAS, 227, 1

\vspace{3 mm}

\noindent
Lacey C., Cole S., 1993, MNRAS, 262, 627

\vspace{3 mm}

\noindent
Limber, D. N. 1953, ApJ, 117, 134

\vspace{3 mm}

\noindent
Loveday J., Peterson B. A., Maddox S. J., Efstathiou G., 1996, ApJS,
 107, 201

\vspace{3 mm}

\noindent
Matsubara T., Suto Y., 1996, ApJL, 470 L1+

\vspace{3 mm}

\noindent
Matsubara T., Szalay A. S., 2001, ApJL, 556, L67

\vspace{3 mm}

\noindent
Mo H. J., White S. D. M., 1996, MNRAS, 282, 347

\vspace{3 mm}

\noindent
Mountrichas, G., Shanks T., 2007, MNRAS, 380, 113M

\vspace{3 mm}

\noindent
Mountrichas, G., Shanks T., 2007, submitted to MNRAS, astro-ph 0712.3255

\vspace{3 mm}

\noindent
Myers A. D., Outram P. J., Shanks T., Boyle B. J., Croom S. M., Loaring
 N. S., Miller L., Smith R. J., 2003, MNRAS, 342, 467

\vspace{3 mm}

\noindent
Myers A. D., Outram P. J., Shanks T., Boyle B. J., Croom S. M., Loaring
 N. S., Miller L., Smith R. J., 2005, MNRAS, 359, 741

\vspace{3 mm}

\noindent
Myers A.D., Brunner R.J., Richards G.T., Nichol R.C., Schneider D.P., Vanden Berk D.E., Scranton R., Gray A.G., Brinkmann J., 2006, ApJ, 638, 622

\vspace{3 mm}

\noindent
Myers A. D., Brunner R. J., Richards G. T., Nichol R. C., Schneider D. P., Bahcall N. A., 2007, ApJ, 658, 99M

\vspace{3 mm}

\noindent
Navarro J. F., Frenk C. S., White S. D. M., 1997, ApJ, 490, 493

\vspace{3 mm}

\noindent
Padmanabhan, N., White, M., Eisenstein, D. J., 2006, American
 Astronomical Society Meeting 209, Vol. 38, p.1063

\vspace{3 mm}

\noindent
Peacock J. A., et al., 2001, Nature, 410, 169 

\vspace{3 mm}

\noindent
Peebles P. J. E., 1980, The large-scale structure of the universe.
 Researh supported by the National Science Foundation. Princeton, N. J.,
 Princeton University Press, 1980, 435p

\vspace{3 mm}

\noindent
Peebles P. J. E., 1984, ApJ, 284, 439

\vspace{3 mm}

\noindent
Phillipps, S., Fong, R., Fall, R. S., Ellis S. M., MacGillivray, H. T., 1978, MNRAS, 182, 673 

\vspace{3 mm}

\noindent
Porciani C., Magliocchetti M., Norberg P. 2004, MNRAS, 355, 1010

\vspace{3 mm}

\noindent
Porciani C., Norberg P., 2006, MNRAS, 371, 1824

\vspace{3 mm}

\noindent
Ratcliffe, A., Shanks, T., Broadbent, A., Parker, Q. A., Watson, F. G.,
 Oates, A. P., Fong, R., Collins, C. A., 1996, MNRAS, 281L, 47R

\vspace{3 mm}

\noindent
Ratcliffe, A., Shanks, T., Parker, Q. A., Fong, R., 1998, MNRAS, 296,
 191

\vspace{3 mm}

\noindent
Richards G. T., et al., 2005, MNRAS, 360, 839

\vspace{3 mm}

\noindent
Ross, N. P., et al., 2007, MNRAS, 381, 573

\vspace{3 mm}

\noindent
Ross, N. P., Shanks, T., Cannon, R. D., Wake, D. A., Sharp, R. G.,
 Croom, S. M., Peacock, John A., astro-ph/0704.3739, submitted to MNRAS

\vspace{3 mm}

\noindent
Rubin, V. C., 1954, Proc. N.A.S., 40, 541

\vspace{3 mm}

\noindent
Saunders, W., Rowan-Robinson, M., Lawrence, A., 1992, MNRAS, 258, 134

\vspace{3 mm}

\noindent
Sheth R. K., Mo H. J., Tormen G., 2001, MNRAS, 323, 1

\vspace{3 mm}

\noindent
Smith R. J. Boyle B. J., Shanks T., Croom S. M., Miller L., Read M.,
 1997, IAUS, 179, 348

\vspace{3 mm}

\noindent
Wake D. A., et al., 2004, ApJL, 610, L85

\clearpage

\clearpage

\clearpage

\begin{table*}
\caption{$r_0$ and $\gamma$ values from the fits on the QSO-2SLAQ (photometric) LRG $\xi (r)$ measurements.}
\centering
\setlength{\tabcolsep}{1.5mm}
\begin{tabular}{lccc}
 \multicolumn{4}{c}{$\xi (r)$}\\
       \hline
$$ & \multicolumn{3}{c}{photometric 2SLAQ LRGs} \\
       \hline
$$ & 2SLAQ QSOs& 2QZ QSOs& SDSS QSOs \\
       \hline \hline
$r_0$ & $7.5_{-0.3}^{+0.3}$ & $8.0_{-0.4}^{+0.4}$ & $7.0_{-0.3}^{+0.3}$ \\
       \hline \hline
$-\gamma$ & $1.7_{-0.2}^{+0.2}$ & $1.7_{-0.2}^{+0.2}$ & $1.8_{-0.1}^{+0.1}$ \\
       \hline
\label{table:photometric}
\end{tabular}
\end{table*}

\begin{table*}
\caption{$s_0$ and $\gamma$ values from the fits on the QSO-2SLAQ (spectroscopic) LRG $\xi (s)$ measurements, on scales of 5-25h$^{-1}$Mpc.}
\centering
\setlength{\tabcolsep}{1.5mm}
\begin{tabular}{lccc}
 \multicolumn{4}{c}{$\xi (s)$}\\
       \hline
$$ & \multicolumn{3}{c}{spectroscopic 2SLAQ LRGs} \\
       \hline
$$ & 2SLAQ QSOs& 2QZ QSOs& SDSS QSOs\\
       \hline \hline
$s_0$ & $8.2_{-0.1}^{+0.1}$ & $8.0_{-0.1}^{+0.3}$ & $7.5_{-0.2}^{+0.3}$  \\
       \hline \hline
$-\gamma$ & $1.6_{-0.1}^{+0.2}$  & $1.7_{-0.1}^{+0.2}$ & $1.8_{-0.3}^{+0.2}$    \\
       \hline
\label{table:spectroscopic}
\end{tabular}
\end{table*}

\begin{table*}
\caption{$r_0$ and $\gamma$ values from the fits on the
 $w_p(\sigma)/\sigma$ measurements, on scales of 5-25h$^{-1}$Mpc.}
\centering
\setlength{\tabcolsep}{1.5mm}
\begin{tabular}{lccc}
       \hline
 \multicolumn{4}{c}{$w_p(\sigma)/\sigma$}\\
       \hline
$$ & \multicolumn{3}{c}{spectroscopic 2SLAQ LRGs} \\
       \hline
$$ & 2SLAQ QSOs& 2QZ QSOs& SDSS QSOs \\
       \hline \hline
$r_0$ & $6.8_{-0.3}^{+0.1}$ & $6.0_{-0.2}^{+0.4}$ & $6.3_{-0.1}^{+0.3}$\\
       \hline \hline
$-\gamma$ & $1.7_{-0.3}^{+0.2}$  & $1.5_{-0.2}^{+0.1}$ & $1.8_{-0.1}^{+0.1}$\\
       \hline
\label{fig:table_wp}
\end{tabular}
\end{table*}

\begin{table*}
\caption{$r_0$ and $\gamma$ values from the fits on the $\xi (r)$
 measurements, on scales of 5-25h$^{-1}$Mpc.}
\centering
\setlength{\tabcolsep}{1.5mm}
\begin{tabular}{lccc}
       \hline
 \multicolumn{4}{c}{$\xi (r)$}\\
       \hline
$$ & \multicolumn{3}{c}{spectroscopic 2SLAQ LRGs} \\
       \hline
$$ & 2SLAQ QSOs& 2QZ QSOs& SDSS QSOs \\
       \hline \hline
$r_0$ & $7.0_{-0.1}^{+0.2}$ & $7.0_{-0.3}^{+0.3}$ & $5.5_{-0.2}^{+0.4}$ \\
       \hline \hline
$-\gamma$ & $2.1_{-0.1}^{+0.2}$  & $1.6_{-0.1}^{+0.1}$ & $2.3_{-0.3}^{+0.2}$\\
       \hline
\label{fig:table_xir}
\end{tabular}
\end{table*}

\begin{table*}
\caption{QSO $b_Q$, $\beta _Q$ and $\langle \omega_z^2\rangle ^{1/2}$ measurements from modelling the redshift-space distortions.}
\centering
\setlength{\tabcolsep}{1.5mm}

\begin{tabular}{lcccccc}
       \hline
 \multicolumn{7}{c}{QSO $b_Q$ and $\beta _Q$}\\
       \hline
$$ & \multicolumn{3}{c}{model I} & \multicolumn{3}{c}{model II} \\

$$ & 2SLAQ QSOs& 2QZ QSOs& SDSS QSOs& 2SLAQ QSOs& 2QZ QSOs& SDSS QSOs \\
       \hline \hline
$b_Q(\simeq\frac{\Omega _m^{0.6}}{\beta _Q})$  & $1.66_{-0.69}^{+1.58}$ & $1.36_{-0.67}^{+1.41}$ & $1.25_{-0.46}^{+0.88}$ & $1.87_{-0.83}^{+2.28}$ & $1.25_{-0.37}^{+1.74}$ & $1.15_{-0.32}^{+1.11}$  \\
       \hline \hline
$\beta _Q$ & $0.45_{-0.22}^{+0.32}$  & $0.55_{-0.38}^{+0.53}$ & $0.60_{-0.25}^{+0.35}$ & $0.40_{-0.22}^{+0.32}$ & $0.60_{-0.35}^{+0.25}$ & $0.65_{-0.37}^{+0.25}$ \\
       \hline \hline
$\langle \omega_z^2\rangle ^{1/2}$ (kms$^{-1})$ & $630$ & $560$ & $670$ & $750$ & $720$ & $710$\\
       \hline
\label{Table:beta_b_redz}
\end{tabular}
\end{table*}

\begin{table*}
\caption{QSO-LRG $\xi _{20}$ cross-correlation measurements, as well as QSO $b_Q$ and $\beta _Q$ measurements, assuming $b_{L(AA\Omega)}=2.35\pm0.20$ and $b_{L(2SLAQ)}=1.90\pm0.08$, from the amplitude results. For consistency, the $\xi _{20}$ measurements for the spectroscopic samples come from $w_p(\sigma)$ and from $\xi (r)$ via $w(\theta)$ for the photometric cases.}
\centering
\setlength{\tabcolsep}{1.5mm}
\begin{tabular}{lccccccccc}
       \hline
 \multicolumn{10}{c}{QSO $b_Q$ and $\beta _Q$}\\
       \hline
$$ & \multicolumn{3}{c}{spectroscopic 2SLAQ LRGs} & \multicolumn{3}{c}{photometric 2SLAQ LRGs}  &
 \multicolumn{3}{c}{photometric AAOmega LRGs} \\
       \hline
$$ & 2SLAQ QSOs& 2QZ QSOs& SDSS QSOs& 2SLAQ QSOs& 2QZ QSOs& SDSS QSOs&2SLAQ QSOs& 2QZ QSOs& SDSS QSOs
 \\
       \hline \hline
$\xi _{20}$ & $0.37\pm0.06$ & $0.33\pm0.10$ & $0.31\pm0.14$ & $0.44\pm0.08$ & $0.49\pm0.06$ & $0.38\pm0.08$ & $0.55\pm0.11$ & $0.50\pm0.11$ & $0.33\pm0.08$\\
       \hline \hline
$b_Q$  & $1.62_{-0.17}^{+0.17}$ & $1.44_{-0.26}^{+0.26}$ &
 $1.37_{-0.37}^{+0.37}$ & $1.91_{-0.21}^{+0.21}$ & $2.15_{-0.16}^{+0.16}$ &$1.67_{-0.18}^{+0.18}$ & $2.17_{-0.28}^{+0.28}$ & $1.95_{-0.27}^{+0.27}$ &
 $1.30_{-0.19}^{+0.19}$ \\
       \hline \hline
$\beta _Q (\simeq\frac{\Omega _m^{0.6}}{b})$ & $0.46_{-0.05}^{+0.05}$  &
 $0.51_{-0.09}^{+0.09}$ & $0.54_{-0.14}^{+0.14}$ & $0.28_{-0.03}^{+0.03}$ & $0.35_{-0.02}^{+0.02}$ & $0.45_{-0.05}^{+0.05}$ & $0.37_{-0.05}^{+0.05}$
 & $0.41_{-0.06}^{+0.06}$ & $0.61_{-0.10}^{+0.10}$   \\
       \hline \hline
$M_{b_J}$ & $-21.7$ & $-22.9$ & $-23.7$ & $-21.7$ & $-22.9$ & $-23.7$ & $-22.2$ & $-23.5$ & $-24.5$\\
        \hline
\label{table:biases_qso}
\end{tabular}
\end{table*}

\end{document}